\documentclass[preprint,twocolumn]{aastex631}
\usepackage{amsmath}
\usepackage{graphicx}
\usepackage{float}

\submitjournal{\apj}

\accepted{May 5, 2025}

\shorttitle{New Mechanism to Support a Supercritical Filament against Radial Collapse}
\shortauthors{Abe et al.}

\begin{document}

\title{Growth of Massive Molecular Cloud Filament by Accretion Flows. II. \\
New Mechanism to Support a Supercritical Filament against Radial Collapse}

\email{abe.daisei@astr.tohoku.ac.jp}

\author[0000-0001-6891-2995]{Daisei Abe}
\affiliation{Astronomical Institute, Tohoku University, Sendai, Miyagi, 980-8578, Japan}

\author[0000-0002-7935-8771]{Tsuyoshi Inoue}
\affiliation{Department of Physics, Faculty of Science
and Engineering, Konan University, Okamoto 8-9-1, Higashinada-ku, Kobe 658-8501, Japan}
\affiliation{Department of Physics, Graduate School of Science, Nagoya University, Furo-cho, Chikusa-ku, Nagoya 464-8602, Japan}

\author[0000-0003-4366-6518]{Shu-ichiro Inutsuka}
\affiliation{Department of Physics, Graduate School of Science, Nagoya University, Furo-cho, Chikusa-ku, Nagoya 464-8602, Japan}

\author[0000-0002-1959-7201]{Doris Arzoumanian}
\affiliation{The Institute for Advanced Study, Kyushu University, Japan}
\affiliation{Department of Earth and Planetary Sciences, Faculty of Science, Kyushu University, Nishi-ku, Fukuoka 819-0395, Japan}
\affiliation{National Astronomical Observatory of Japan, Osawa 2-21-1, Mitaka, Tokyo 181-8588, Japan}

\graphicspath{{./}{figures/}}

\begin{abstract}
{Observations indicate that dense molecular filamentary clouds are sites of star formation.
The filament width determines the {most unstable scale for self-gravitational fragmentation} and influences the stellar mass.
Therefore, {constraining} the evolution of filaments and the origin of their properties {are} important for understanding star formation.
Although {some} observations show a universal width of 0.1 pc, {many} theoretical studies predict the contraction of thermally supercritical filaments ($>$ 17 $M_{\odot}$ pc$^{-1}$) due to radial collapse.
Through non-ideal magnetohydrodynamics simulations with ambipolar diffusion, we explore the formation and evolution of filaments via slow-shock instability at the front of accretion flows.
We reveal that ambipolar diffusion allows the gas in the filament to flow across the magnetic fields {around the shock front}, forming dense blobs behind the concave points of the shock front. 
The blobs transfer momentum that drives internal turbulence.
We name this mechanism the ``\textit{STORM}" (Slow-shock-mediated Turbulent flOw Reinforced by Magnetic diffusion).
The persistence and efficiency of the turbulence inside the filament are driven by the magnetic field and the ambipolar diffusion effect, respectively.
The {STORM} mechanism sustains the width even when the filament reaches very large line masses ($\sim$ 100 $M_{\odot}$ pc$^{-1}$).
}
\end{abstract}

\keywords{stars: formation --- ISM: clouds --- magnetohydrodynamics (MHD)}

\section{Introduction} \label{sec:intro}
The dense filamentary structures found within molecular clouds are sites of star formation~\citep[e.g.,][]{andre2010A&A...518L.102A, andre2014prpl.conf...27A, Hacar2023ASPC..534..153H, Pineda2023ASPC..534..233P}.
According to findings from the \textit{Herschel} Gould Belt survey, star formation occurs within filaments{, defined by dense elongated structures in a molecular cloud,} with line masses exceeding the critical threshold at which gravity dominates over thermal pressure.
This threshold is given by $M_{\mathrm{line,cr,th}} = 2c_{\mathrm{s}}^2 /G \simeq 17\ \mathrm{M_{\odot}\ pc^{-1}},$ where, $c_{\mathrm{s}} \simeq 0.2\ \mathrm{km\ s^{-1}}$ is the isothermal sound speed in typical molecular clouds with a temperature of 10 K, and $G$ denotes the gravitational constant \citep[e.g.,][]{Stodolkiewicz1963, Ostriker1964, InutsukaMiyama1992, Inutsuka1997}.
Various studies have explored the formation mechanisms of molecular filaments~\citep[e.g., ][]{Tomisaka1983,Nagai1998,PadoanNordlund1999ApJ...526..279P,Hennebelle2013,Pudritz2013RSPTA.37120248P,inoue2013ApJ...774L..31I,chenOstriker2014ApJ...785...69C,inutsuka2015A&A...580A..49I,Balfour2017,Federrath2016,abe2021ApJ...916...83A}.
Recently, \citet{abe2021ApJ...916...83A} attempted to provide a comprehensive understanding of filament formation mechanisms.
To this end, they classified the formation mechanisms into five types: Type G (self-gravity in a shock-compressed sheet), C (compressive flows leading to transient filaments), O (oblique shock-induced flows), I (collisions between turbulent sheets), and S (stretching by turbulent shear flows)~\citep[see also the review by][]{Pineda2023ASPC..534..233P}.
In these models, star-forming filaments are created through gas flow along the local magnetic field within a shock-compressed sheet.

As evidenced in several recent works, accretion plays a crucial role in filament evolution.
Molecular line-emission observations provide evidence of perpendicular accretion onto filaments~\citep{Palmeirim2013, shimajiri2019A&A...632A..83S, Chen2020}.
In particular, \citet{shimajiri2019A&A...632A..83S} identified gas accretion onto filaments within a shocked sheet.
{From the theoretical side, a lot of studies show that accretion controls filament evolution~\citep{Heitsch2013ApJ...776...62H, Heitsch2013bApJ...769..115H, HennebelleAndre2013A&A...560A..68H, Clarke2016MNRAS.458..319C, Clarke2017MNRAS.468.2489C, Heigl2020MNRAS.495..758H}.}
The results of the {hydrodynamic} simulation of \citet{Clarke2016MNRAS.458..319C} demonstrated that the most unstable length scale for self-gravitational fragmentation depends on the accretion rate onto the filament.
\citet{HennebelleAndre2013A&A...560A..68H} developed an analytical model of self-gravitating and accreting filaments, which considers the turbulence induced by accretion and its dissipation through ion-neutral friction.

\citet{Arzoumanian2011A&A...529L...6A, Arzoumanian2019A&A...621A..42A}, \citet{Juvela2012A&A...541A..12J}, \citet{KochRosolowsky2015MNRAS.452.3435K}, and \citet{Shimajiri2023A&A...672A.133S} revealed a characteristic width of $\sim$ 0.1 pc {for filaments in the nearby Gould Belt clouds observed with \textit{Herschel} ~\citep[see also][for C$^{18}$O line-emission observations]{Orkisz2019A&A...624A.113O, Suri2019A&A...623A.142S}.}
Remarkably, this width is maintained even for filaments with high-line masses ($\gtrsim$~100~$M_{\rm \odot}$~pc$^{-1}$).
{A filament in NGC6334, which has 
$M_{\mathrm{line}}$~$\sim$ 500--1000 $M_{\odot}$ pc$^{-1}$, also has a width of $\sim$ 0.1 pc~\citep{andre2016A&A...592A..54A}.}
If thermal support alone counteracts gravity, such a high-line mass structure cannot maintain a radial width of 0.1~pc because the gravitationally unstable filaments radially collapse~\citep[e.g.,][]{FischeraMartin2012A&A...542A..77F}.
Some studies have questioned the universality of 0.1~pc filament widths.
For example, \citet{Ossenkopf-Okada2019A&A...621A...5O} found no such characteristic scale in a wavelet decomposition of \textit{Herschel} survey data.
\citet{Panopoulou2022A&A...657L..13P, Panopoulou2022A&A...663C...1P} suggested that filament width depends on distance.
However, \citet{Andre2022A&A...667L...1A} performed a convergence test on the B211/213 filament in Taurus and determined its width as $\sim$~0.1~pc.
{They emphasized that the apparent dependence on distance or spatial resolution results from a beam convolution effect in a structured, turbulent medium for distant (not resolved filaments) and that the width measurements of $\sim$ 0.1 pc are robust for nearby (well resolved) high-contrast filaments.}
Several theoretical studies on the effects of turbulence and/or the magnetic field have also shown a width of 0.1 pc for sub-critical and mildly super-critical filaments \citep[][]{FischeraMartin2012A&A...542A..77F, Auddy2016ApJ...831...46A, PriestleyWhitworth2022MNRAS.512.1407P, Federrath2016, Clarke2020MNRAS.497.4390C}.
Although {\citet{HennebelleAndre2013A&A...560A..68H} and \citet{Heitsch2013ApJ...776...62H} proposed that accretion-driven turbulence can delay the radial collapse {for supercritical filaments}, \citet{Heigl2020MNRAS.495..758H} showed that the unmagnetized accretion-driven turbulence does not provide a supporting pressure force capable of opposing the radial collapse.
The reason why thermally supercritical massive filaments remain wide and do not shrink has lacked a physical explanation.}

\begin{figure*}[t]
\epsscale{1.0}
\plotone{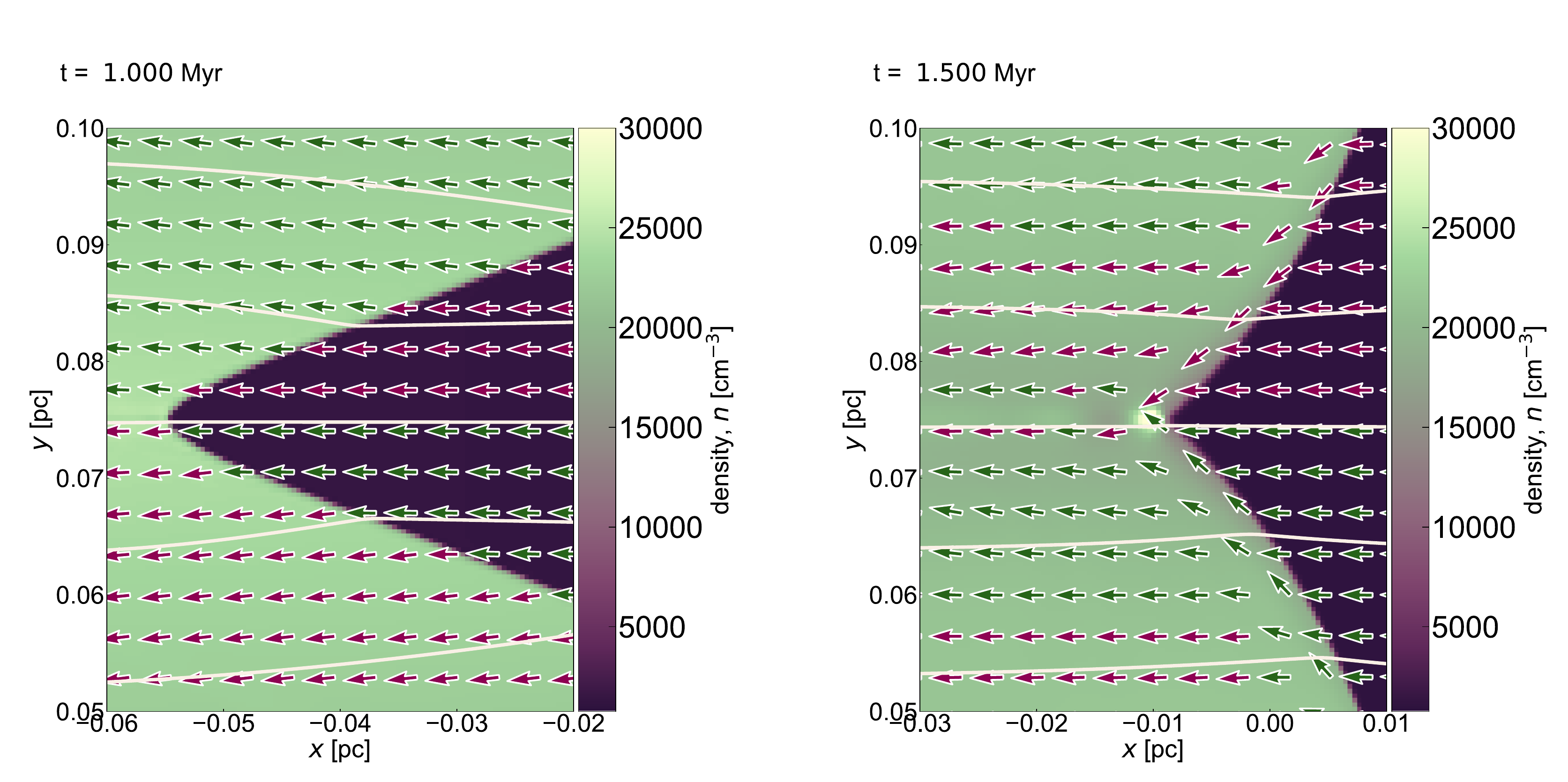}
\caption{\small{{Blow-up} density maps showing the concave part of a single slow-shock surface in (left) the ideal MHD case and (right) MHD with ambipolar diffusion. {We initially set the steady state solution of the shock wave which is located at $x$ = 0 pc and put a perturbation of the shock position in the $y$-direction.} White lines represent the magnetic field lines. The green and pink arrows represent the normalized upward and downward velocity fields, respectively. {We can observe gas flow along the shock front and accumulation in the concave part due to ambipolar diffusion in the right panel in contrast to the left.} \label{fig:singleShock}
}}
\end{figure*}

It should be mentioned that {the} filament width is a crucial parameter for star formation.
\citet{lada2010ApJ...724..687L}, \citet{Heiderman2010ApJ...723.1019H}, \citet{Evans2014ApJ...782..114E}, and \citet{Konyves2015A&A...584A..91K} claimed that star formation in molecular clouds is triggered when the column density reaches $\sim$~10$^{22}$~cm$^{-2}$.
Although this result is apparently unrelated to the filament paradigm of star formation, the critical line mass divided by the width scale of 0.1 pc gives a column density of 10$^{22}$~cm$^{-2}$, suggesting that the filament width is a key length scale connecting the column density threshold of star formation to the critical line mass of star-forming filaments~\citep{andre2014prpl.conf...27A}.
{According to the linear theory, the most unstable mode (i.e., the fragmentation length scale) for the self-gravitational fragmentation of a cylindrical filament depends on the filament width~\citep{Stodolkiewicz1963, InutsukaMiyama1992, Hanawa2019ApJ...881...97H}, indicating that the characteristic core mass should be determined by the filament width and line mass.
However, it should be mentioned that actual fragmentation occurs sensitively depending on the initial amplitude of line-mass fluctuation of the filament~\citep{Inutsuka1997}.
Thus, the resulting core mass function is predicted from the initial spectrum of the line mass fluctuations~\citep{Inutsuka2001ApJ...559L.149I}.
\citet{Inutsuka2001ApJ...559L.149I} also have shown that the core mass function follows a Salpeter-like power law for a filament with line mass fluctuation following Kolmogorov-like spectrum.
The fragmentation in the cases of various velocity fluctuations on the filament is theoretically studied by \citet{Clarke2017MNRAS.468.2489C, Clarke2020MNRAS.497.4390C}.
In addition, \citet{Andre2019A&A...629L...4A} found that the observed filament line mass function follows a Salpeter-like power law, suggesting that the core mass function (and by extension the initial mass function of stars) is inherited by the filament line mass function if the filaments fragment at the universal width scale.
In reality, the combination (more precisely, the convolution) of the theory proposed by \citet{Inutsuka2001ApJ...559L.149I} and the filament line mass function might determine the core mass function.
Thus, we expect that the clarification of the physical origin of the uniform filament width should contribute to our understanding of the origin of the initial mass function of stars.}


A shock wave with an Alfv\'{e}n Mach number $\mathcal{M}_{\rm A}<1$ and a sonic Mach number $\mathcal{M}_{\rm s}>1$ is called a \textit{slow-shock}.
Given that massive filaments are formed in the post-shock layer threaded by a strong magnetic field with energy exceeding the kinetic energy of inflowing flows toward the filaments~\citep[the type O/C mechanism described in][]{inoue2013ApJ...774L..31I, Inoue2018PASJ...70S..53I, chenOstriker2014ApJ...785...69C, Chen2020}, such slow-shocks naturally confine the filament surface.
{Although the occurrence of slow-shocks has not yet been definitively confirmed through observations, their presence is a plausible assumption given the MHD nature of the ISM.}
Linear stability analyses \citep{LessenDeshpande1967JPlPh...1..463L, Edelman1989Ap.....31..758E, StoneEdelman1995ApJ...454..182S} have shown that a slow-shock front is corrugationally unstable (slow-shock instability, hereafter SSI).
As a corrugated shock front usually produces turbulent flows behind the shock~\citep[e.g.,][]{Inoue2012ApJ...744...71I, InoueInutsuka2012ApJ...759...35I, Friedmann1922ZPhy...10..377F, Crocco1937ZaMM...17....1C}, an SSI is expected to deposit additional energy in the filament.

The SSI dynamics in molecular clouds can be modified by ambipolar diffusion.
The magnetic Reynolds number of the flow is given as
\begin{equation}
\mathcal{R}_{\mathrm{AD}} = \frac{v\ell}{\eta_{\rm AD}},
\end{equation}
where, $v$ and $\ell$ denote the flow velocity and length scale, respectively, $\eta_{\rm AD}$ denotes the ambipolar diffusion coefficient, calculated as
\begin{equation}
\eta_{\mathrm{AD}} = \frac{ B^2}{4\pi \gamma_{\mathrm{in}} \rho_{\mathrm{n}}\rho_{\mathrm{i}}}.
\label{eqs:ambi coeff intro}
\end{equation}
$B$, $\gamma_{\mathrm{in}} \equiv \langle \sigma_{\mathrm{in}}v_{\rm in} \rangle / (m+m_{\mathrm{i}})$, $\rho_{\mathrm{n}}$, and $\rho_{\mathrm{i}}$ denote the magnetic field strength, the ion-neutral collision rate, the neutral mass density, and the ion mass density, respectively.
$\sigma_{\mathrm{in}}$, $v_{\rm in}$, $m=2.4m_{\mathrm{H}}$, and $m_{\mathrm{i}}=29m_{\mathrm{H}}$ represent the ion-neutral (Langevin) cross-section, the relative velocity between a neutral molecule and ion, mean molecule mass, and mean ion mass, respectively.
{In molecular clouds, since the temperature is low {($\sim$10--15 K)}, the $v_{\rm in}$ is also low enough to induce a dipole moment in the neutral molecule, leading to the effective cross-section with approximately $\gamma_{\mathrm{in}} \propto v_{\rm in}^{-1}$. Then $\gamma_{\mathrm{in}}$ is a constant and its {value $\gamma_{\mathrm{in}}$~=~$3.5\times10^{13}$~$\mathrm{cm^3\ g^{-1}\ s^{-1}}$ (as computed by~\citet{Draine1983ApJ...264..485D}).}}
Assuming that the ionization by cosmic rays balances the recombination of ions, $\rho_{\mathrm{i}}$ can be expressed as $C \rho_{\rm n}^{1/2}$.
In this study, we set $C=3\times 10^{-16}$~cm$^{-3/2}$~g$^{1/2}$~\citep[][]{shu1992pavi.book.....S, HennebelleInutsuka2019FrASS...6....5H}.
To obtain the characteristic length scale below which the effect of ambipolar diffusion becomes non-negligible, we set $\mathcal{R}_{\mathrm{AD}}= 1$ to obtain
\begin{align}
\ell_{\mathrm{AD}}= \eta_{\mathrm{AD}}/v = &0.09\ \mathrm{pc}\ \left(\frac{B}{30\ \mathrm{\mu G}}\right)^{2}\left(\frac{n}{10^3\ \mathrm{cm^{-3}}}\right)^{-3/2} \nonumber \\ 
& \times \left(\frac{v}{1\ \mathrm{km\ s^{-1}}}\right)^{-1}\nonumber  \\
& \times \left( \frac{C}{3\times 10^{-16}\ \rm{cm^{-3/2} g^{1/2}}} \right)^{-1}.
\label{eqs:ambi scale}
\end{align}
In this formulation, the characteristic scale of the ambipolar diffusion is comparable to the observed filament width of 0.1 pc (see above).
The linear evolution of SSI with ambipolar diffusion has been studied analytically by \citet{Bethune2023PhFl...35h4105B} and numerically by \citet{Abe2024ApJ...961..100A}.
In both studies, the most unstable scale of the SSI {is} scaled with $\ell_{\rm AD}$.
{Note that the value of ``0.09 pc" is a rough estimation to evaluate whether the ambipolar diffusion works in 0.1 pc wide filaments.}
\citet[][]{SnowHillier2021MNRAS.506.1334S}, who performed two-dimensional (2D) two-fluid simulations of SSI in partially ionized gas in the solar chromosphere, demonstrated that neutral fluid stabilizes the SSI on a small scale.
They also found new features such as gas accumulation in valleys of shock front corrugations.
{Figure \ref{fig:singleShock} shows snapshots of the density field in single shock models for ideal (the left panel) and non-ideal {MHD} including ambipolar diffusion (the right panel) cases. The upstream density, velocity, and strength of the magnetic field are 1000 cm$^{-3}$, 1 km s$^{-1}$, and 30 $\rm{\mu}$G, respectively.
We can observe the gas accumulation in valleys of shock front corrugations in the non-ideal MHD model.}
However, these previous studies considered the simple case of single-shock propagation.
In reality, the filaments in O- and C-type mechanisms, form and grow through accretion flows causing \textit{two} slow-shocks.
Thus, we must investigate whether two slow-shocks drive turbulence.

In this paper, we perform non-ideal MHD simulations of turbulence generation within filaments and determine whether the energy supply of turbulence can support the filament against gravitational collapse.
The paper is organized as follows: \S \ref{sec:setup} describes the setup of our simulations.
In \S \ref{sec:Results}, we show the mechanism of turbulence generation and demonstrate that a turbulent massive filament can maintain the width as observed.
A simple analytic model of massive filament width is developed in \S \ref{sec: Discussion}.
\S \ref{sec: summary} summarizes the paper.

\section{Simulation Setup} \label{sec:setup}

\begin{figure*}[t!]
\epsscale{0.9}
\plotone{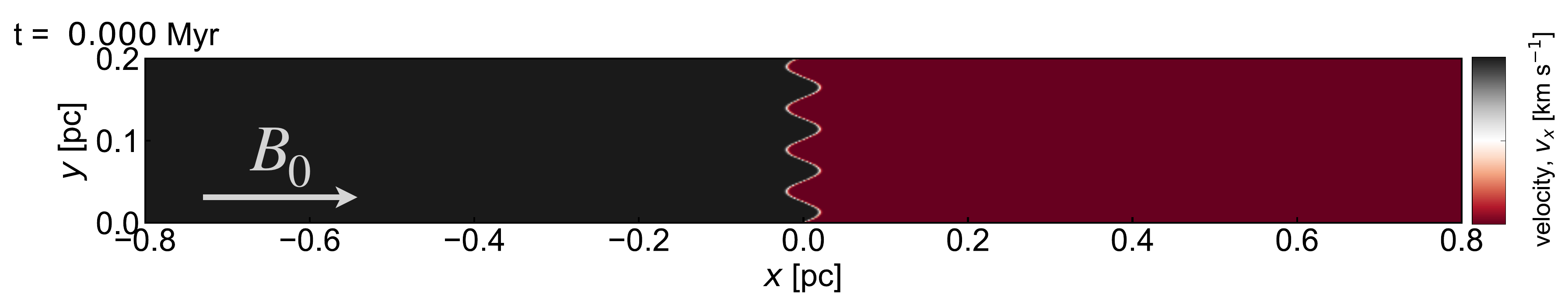}
\caption{An example of the initial conditions for the 2D simulations (model SingleModeIdeal): The color bar represents the x-component of velocity. The initial density, the upstream velocity, the strength of the magnetic field, {the wavelength and amplitude of the shock position disturbance} are 1000 cm$^{-3}$, 1 km s$^{-1}$, 30 $\mu$G, 0.05pc, and 0.02 pc, respectively.\label{fig:inicon2d}
}
\end{figure*}

\begin{table*}
\caption{Model parameters of the 2D simulations presented in this study \label{tab:Model parameters to study the nonlinear evolution.}}
 \centering
  \begin{tabular}{ccccccc}
   \hline
  Model Name   & 2$\pi k^{-1}_{\rm y}$ [pc] & Ambipolar Diffusion & $n_0$ [cm$^{-3}$] & $B_0$ [$\mu$G] & $v_{x0}$ [km s$^{-1}$]& $C$ [cm$^{-3/2}$ g$^{1/2}$] \\
   \hline \hline
    SingleModeIdeal                   & 0.05  & No  & 1,000 & 30 & 1.0 & - \\
    SingleModeAD                      & 0.05  & Yes & 1,000 & 30 & 1.0 & 3 $\times$ 10$^{-16}$ \\
    RandIdeal                         & $L_{\mathrm{box,y}}/l$     & No  & 1,000 & 30 & 1.0 & - \\
    RandAD (fiducial)                 & $L_{\mathrm{box,y}}/l$     & Yes & 1,000 & 30 & 1.0 & 3 $\times$ 10$^{-16}$ \\
    RandHD                            & $L_{\mathrm{box,y}}/l$     & -   & 1,000 & -  & 1.0 & - \\
    RandAD-hd                         & $L_{\mathrm{box,y}}/l$     & Yes & 1,600 & 30 & 1.0 & 3 $\times$ 10$^{-16}$ \\
    RandAD-ld                         & $L_{\mathrm{box,y}}/l$     & Yes & 800   & 30 & 1.0 & 3 $\times$ 10$^{-16}$ \\
    RandAD-sb                         & $L_{\mathrm{box,y}}/l$     & Yes & 1,000 & 50 & 1.0 & 3 $\times$ 10$^{-16}$ \\
    RandAD-wb                         & $L_{\mathrm{box,y}}/l$     & Yes & 1,000 & 24 & 1.0 & 3 $\times$ 10$^{-16}$ \\
    RandAD-hv                         & $L_{\mathrm{box,y}}/l$     & Yes & 1,000 & 30 & 1.3 & 3 $\times$ 10$^{-16}$ \\
    RandAD-lv                         & $L_{\mathrm{box,y}}/l$     & Yes & 1,000 & 30 & 0.8 & 3 $\times$ 10$^{-16}$ \\
    RandAD-lc                   & $L_{\mathrm{box,y}}/l$  & Yes & 1,000 & 30 & 1.0 & 3 $\times$ 10$^{-17}$ \\
    RandAD-hc                   & $L_{\mathrm{box,y}}/l$  & Yes & 1,000 & 30 & 1.0 & 3 $\times$ 10$^{-15}$ \\
    RandAD-hc2                  & $L_{\mathrm{box,y}}/l$  & Yes & 1,000 & 30 & 1.0 & 3 $\times$ 10$^{-14}$ \\
    RandAD-hc3                  & $L_{\mathrm{box,y}}/l$  & Yes & 1,000 & 30 & 1.0 & 3 $\times$ 10$^{-13}$ \\
   \hline
  \end{tabular}
\end{table*}


We conduct 2D (\S \ref{subsec:setup2d}) and three-dimensional (3D) (\S \ref{subsec:setup3d}) ideal/non-ideal MHD simulations using the Athena++ code~\citep{stone2020ApJS..249....4S,Tomida2023ApJS..266....7T}.
We apply the second-order-accurate van Leer predictor-corrector scheme and the piecewise linear method to the primitive variables of equation integration.
We also adopted the Roe solver because it is more numerically stable in nonlinear regimes than the HLLD and LHLLD solvers (see Appendix A and B in \citet{Abe2024ApJ...961..100A})\footnote{In \citet{Abe2024ApJ...961..100A}, we introduced the physical viscosity in our simulations to suppress the carbuncle phenomenon and to study the linear growth of SSI with ambipolar diffusion.
Here, the physical viscosity is omitted because the carbuncle phenomenon occurs when the normal direction of a shock aligns with the numerical grid.
Our initial conditions impose a somewhat large initial corrugation on the shock front, which suppresses the carbuncle phenomenon.
As the SSI grows, the shock front corrugation is maintained and the unphysical carbuncle phenomenon never grows sufficiently to dominate the results.
{We {performed} a simulation using the LHLLD solver, which prevents the carbuncle phenomenon, and confirmed that the results are not different from those obtained with the Roe solver (see Appendix \ref{sec: Difference by Solvers}).}}.
{To ensure the divergence-free condition, $\nabla \cdot \boldsymbol{B} = 0$, we use the constrained transport method~\citep{StoneGardiner2009NewA...14..139S}.}
The following equations are solved:
\begin{equation}
\frac{\partial \rho}{\partial t}+\nabla \cdot(\rho \boldsymbol{v})=0,
\label{eqs:eoc}
\end{equation}
\begin{equation}
\frac{\partial \rho \boldsymbol{v}}{\partial t}+\nabla \cdot\left(\rho \boldsymbol{vv}-\frac{\boldsymbol{B B}}{4\pi}+P^{*}\boldsymbol{I}\right)=-\rho \nabla \Phi,
\label{eqs:eom}
\end{equation}
\begin{equation}
\begin{aligned}
\frac{\partial E}{\partial t}
+ \nabla \cdot \Bigg[
    \left(E + P^{*}\right) \boldsymbol{v}
    - \boldsymbol{B} (\boldsymbol{B} \cdot \boldsymbol{v}) \\
    + \frac{\eta_{\mathrm{AD}}}{|\boldsymbol{B}|^{2}}
      \left\{ \boldsymbol{B} \times (\boldsymbol{J} \times \boldsymbol{B}) \right\}
      \times \boldsymbol{B}
\Bigg] = 0,
\label{eqs:ee}
\end{aligned}
\end{equation}
\begin{equation}
\frac{\partial \boldsymbol{B}}{\partial t} - \boldsymbol{\nabla} \times \left[(\boldsymbol{v} \times \boldsymbol{B}) - \frac{\eta_{\mathrm{AD}}}{|\boldsymbol{B}|^{2}} \boldsymbol{B} \times(\boldsymbol{J} \times \boldsymbol{B})\right]=0,
\label{eqs:ie}
\end{equation}
\begin{equation}
\nabla^2 \Phi=4 \pi G \rho,
\end{equation}
where $\rho, \boldsymbol{v}, \boldsymbol{B}$ and $\Phi$ represent the density, velocity, magnetic field, and gravitational potential, respectively, $P^{*}=p+B^{2} / (8\pi)$ (with $p$ denoting pressure), and $E=e+ \rho v^{2}/2 + B^{2}/(8\pi)$ (with $e$ denoting thermal energy).
As the isothermal treatment is justified in dense regions of molecular clouds, we set {the ratio of specific heats of} $\gamma = 1.01$.
The term $\boldsymbol{J}=\nabla \times \boldsymbol{B}$ represents the current density.
From Eqs. (\ref{eqs:ambi coeff intro}), the ambipolar diffusion coefficient $\eta_{\mathrm{AD}}$ is given by
\begin{equation}
\eta_{\mathrm{AD}} = \frac{B^2}{4\pi \gamma_{\mathrm{in}} \rho_{\mathrm{n}} \rho_{\mathrm{i}}} = \frac{B^2}{4\pi \gamma_{\mathrm{in}} C \rho^{3/2}}.
\end{equation}
{This formalization accounts for a decrease in the ionization fraction behind the shock because we assume the balance between ionization by cosmic rays and recombination.
Therefore, the ionization degree is not spatially constant in our simulation.}

\begin{figure*}[]
\epsscale{1.}
\plotone{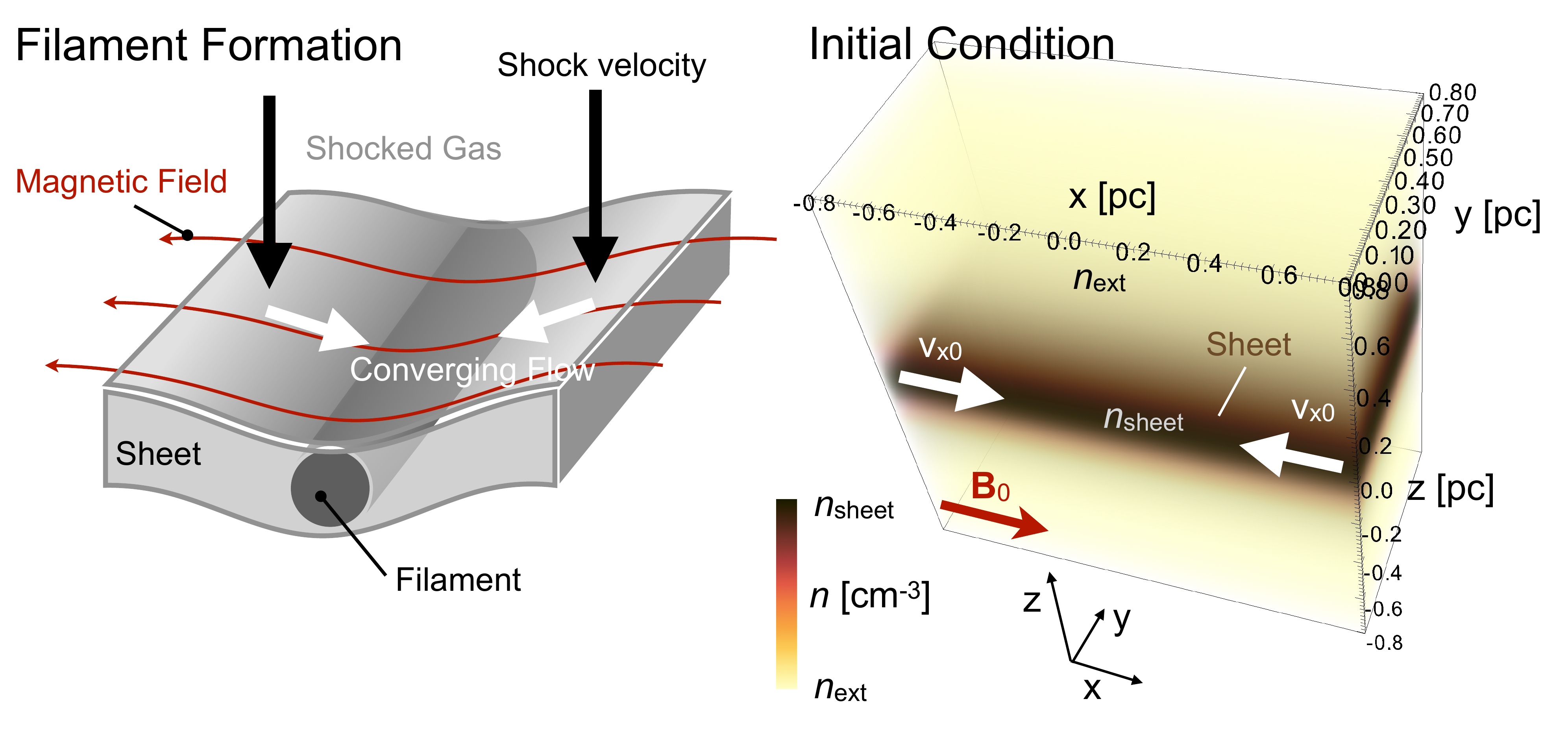}
\caption{Schematic of gas flows at a massive filament formation site (left) and our initial condition of the 3D simulations (right), constructed to resemble the situation in the left panel. The color bar represents the density magnitude. The black and white arrows indicate the orientations of the magnetic field and the converging flows, respectively.\label{fig:inicon}
}
\end{figure*}

\begin{figure*}[]
\epsscale{1.15}
\plotone{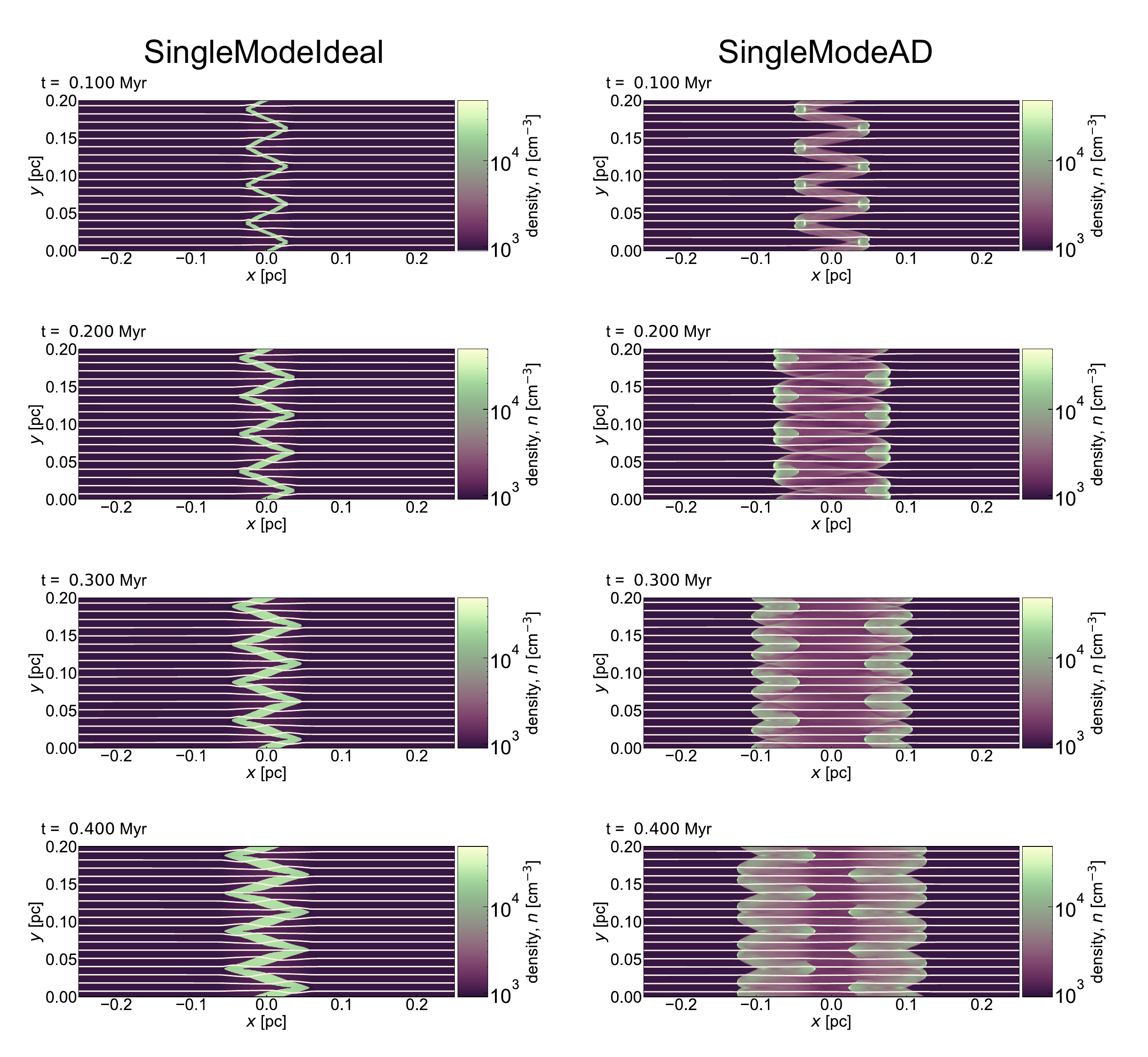}
\caption{\small{Density maps generated by models SingleModeIdeal (left column) and SingleModeAD (right column) at times $t$ = 0.1, 0.2, 0.3, and 0.4 Myr (top to bottom). The white lines represent the magnetic field lines. {In both models, the initial density, the upstream velocity, the strength of the magnetic field, and the length scale of shock position disturbance are 1000 cm$^{-3}$, 1 km s$^{-1}$, 30 $\mu$G, and 0.05pc, respectively.} {Compared to the ideal MHD case, the shock-compressed layer {filament} promptly expands in the non-ideal MHD case.}\label{fig:2ddensity005}
}}
\end{figure*}

\begin{figure*}[!t]
\epsscale{1.1}
\plotone{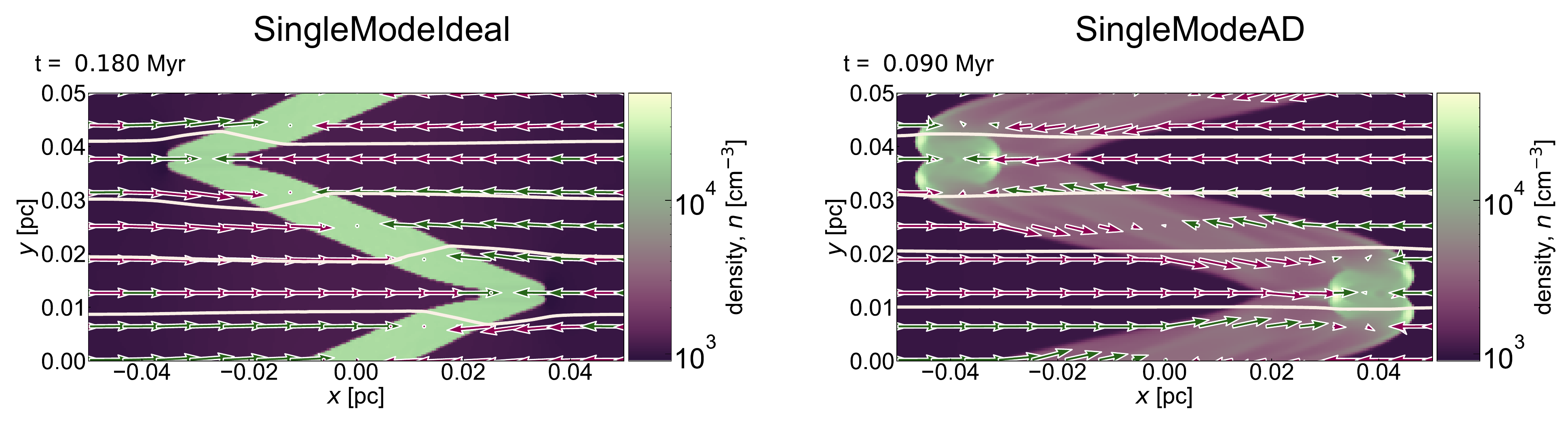}
\caption{\small{Density fields and velocity structures obtained by models SingleModeIdeal (left) and SingleModeAD (right). White lines represent magnetic field lines. Arrows with green and pink represent velocity vectors with {upward ($v_{\rm y} > 0$) and downward velocities ($v_{\rm y} < 0$)}, respectively. {Most of the converging flow gas stops at the shock fronts in the ideal MHD case but the gas moves through the post-shock region in the non-ideal MHD case.} \label{fig:twoShock}
}}
\end{figure*}
\begin{figure*}[!t]
\epsscale{1.15}
\plotone{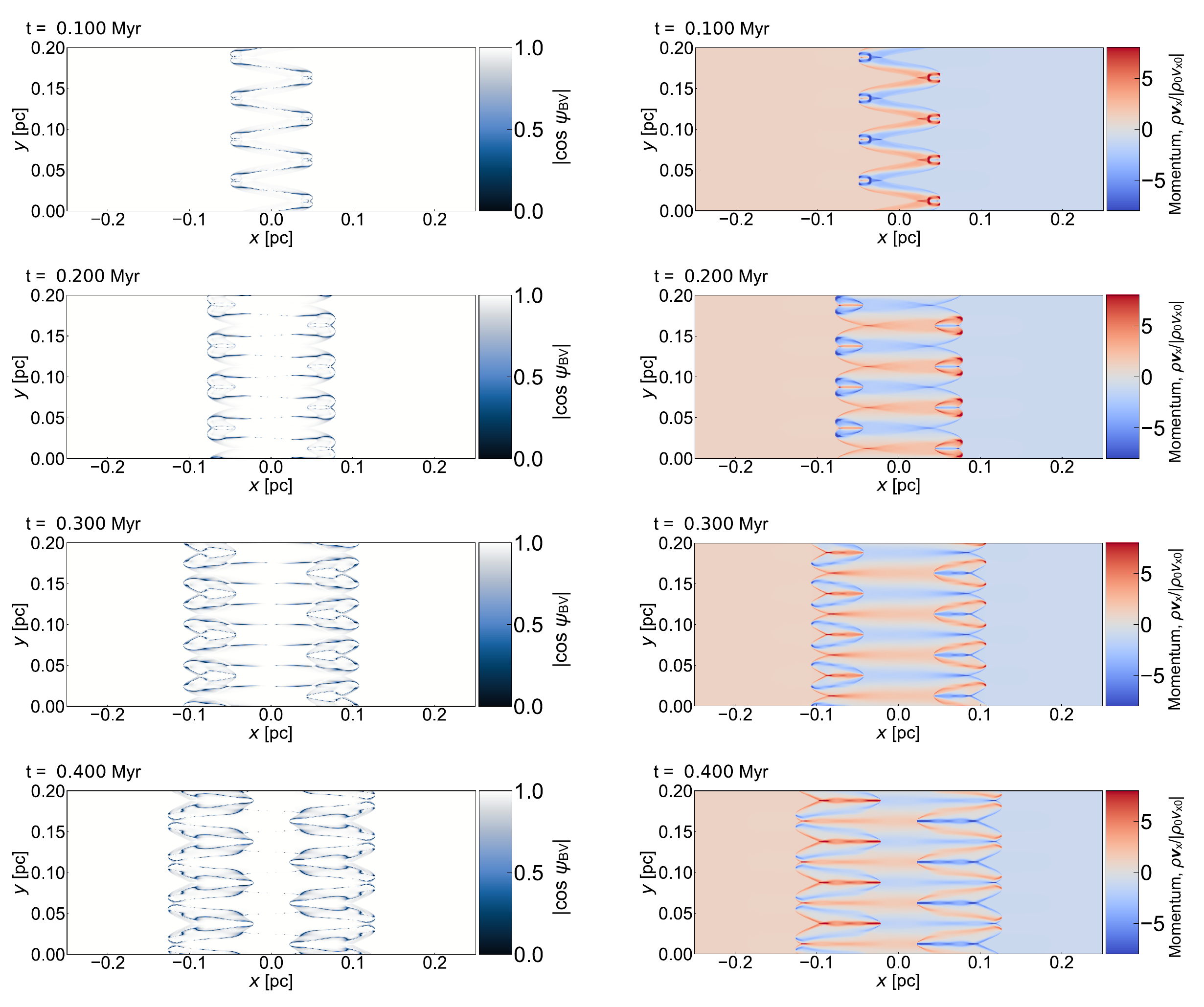}
\caption{\small{
{
\textit{Left:} $|\cos \Psi_{\mathrm{BV}}|$ maps {for} SingleModeAD at time $t$ = 0.1, 0.2, 0.3, and 0.4 Myr (top to bottom). The direction of the magnetic field is initially parallel or anti-parallel to the velocity {(along the $x$-axis)}.
In the blue region, the direction of the magnetic field and velocity field are misaligned indicating that the gas moves across the magnetic field due to ambipolar diffusion.
We can confirm that ambipolar diffusion impacts the vicinity of the shock fronts.
\textit{Right:} $x$-component of momentum generated by model SingleModeAD at time $t$ = 0.1, 0.2, 0.3, 0.4 Myr (top to bottom).
We can observe that the blobs generated in the concave region in the shock fronts {(for instance [x, y] = [0.025 pc, 0.0125 pc] or [-0.05 pc, 0.0375 pc] at $t$ = 0.1 Myr)} flow toward the center of the shocked layer and stir the gas inside the shock-compressed layer, injecting momentum {(for instance [x, y] = [-0.025 pc, 0.0375 pc] at $t$ = 0.4 Myr). See a schematic illustration in Figure \ref{fig:mechssiad}.}
}
\label{fig:momx}
}}
\end{figure*}

\begin{figure*}[]
\epsscale{1.1}
\plotone{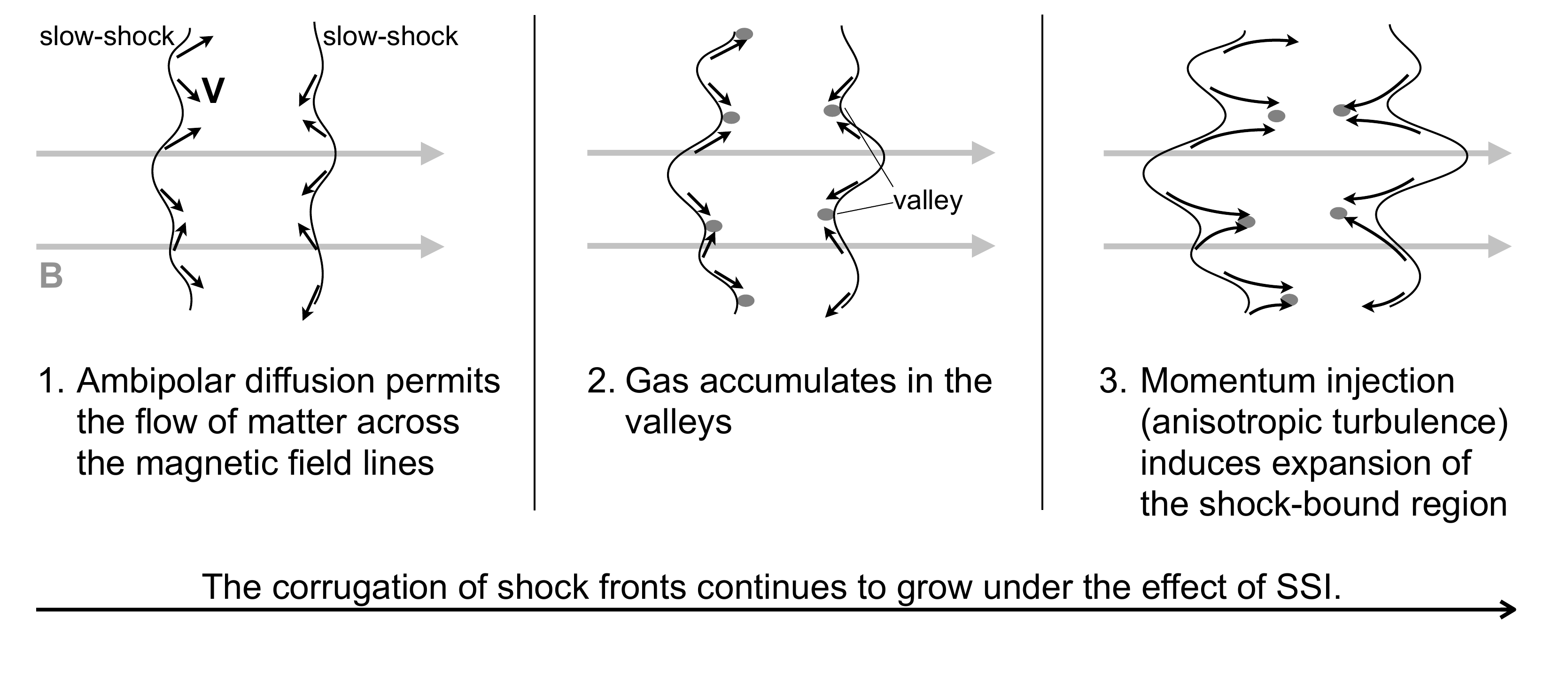}
\caption{\small{Schematic showing the mechanism of fast expansion driven by the nonlinear effect of two-shocks SSI including ambipolar diffusion {without self-gravity}
\label{fig:mechssiad}
}}
\end{figure*}

\begin{figure*}[]
\epsscale{1.2}
\plotone{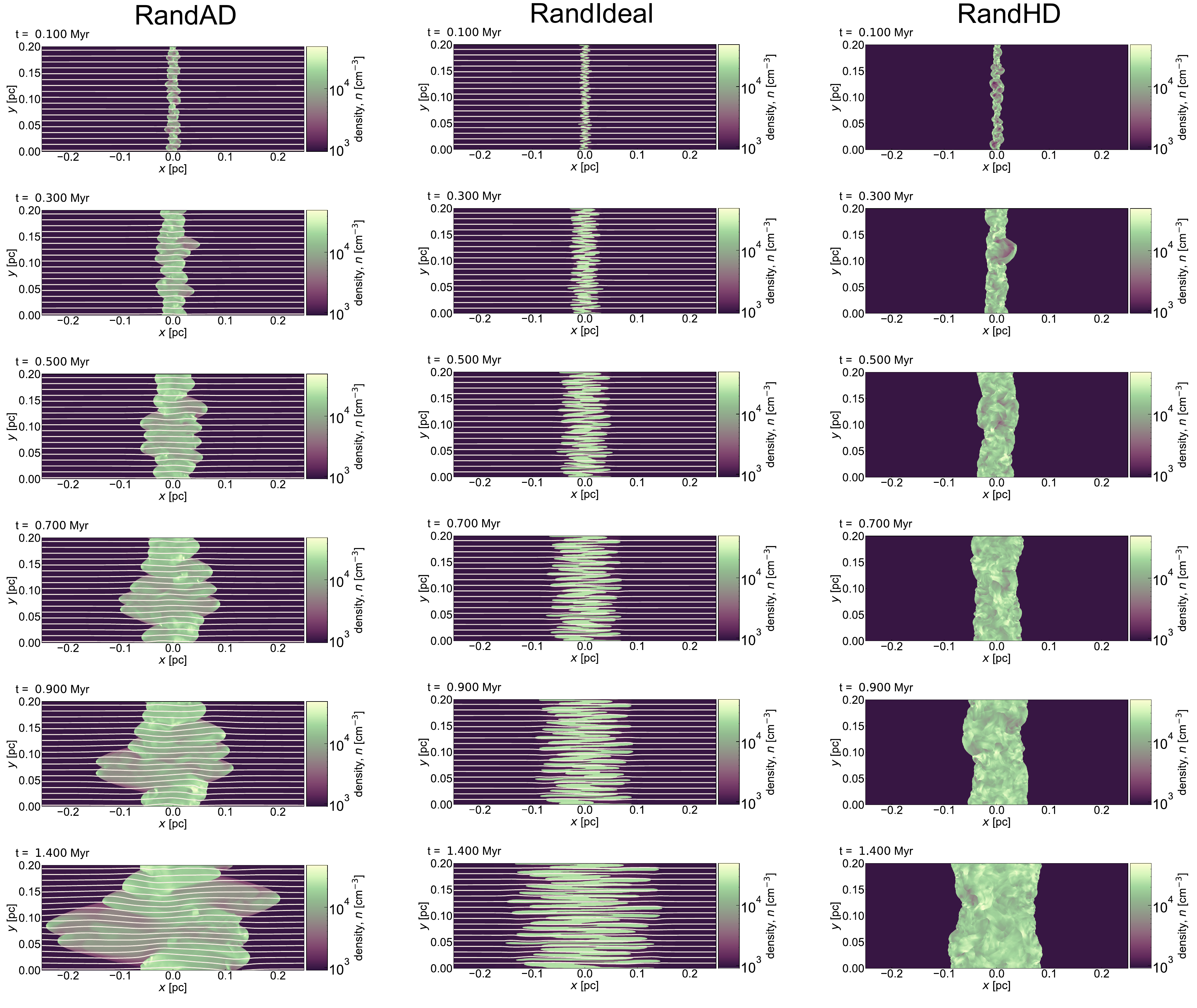}
\caption{\small{Density maps produced by models RandAD, RandIdeal, and RandHD at time $t$ = 0.1, 0.3, 0.5, 0.7, 0.9, and 1.4 Myr (top to bottom). White lines represent the magnetic field lines. {In these models, the initial density and the upstream velocity are 1000 cm$^{-3}$ and 1 km s$^{-1}$, respectively. For models RandAD and RandIdeal, the strength of the magnetic field is 30 $\mu$G. Owing to the {STORM}, the post-shock region expands faster (left panels) than in the case without ambipolar diffusion (center panels) and pure hydro case (right panels).}\label{fig:2ddensityrand}
}}
\end{figure*}

\begin{figure*}[]
\epsscale{1.2}
\plotone{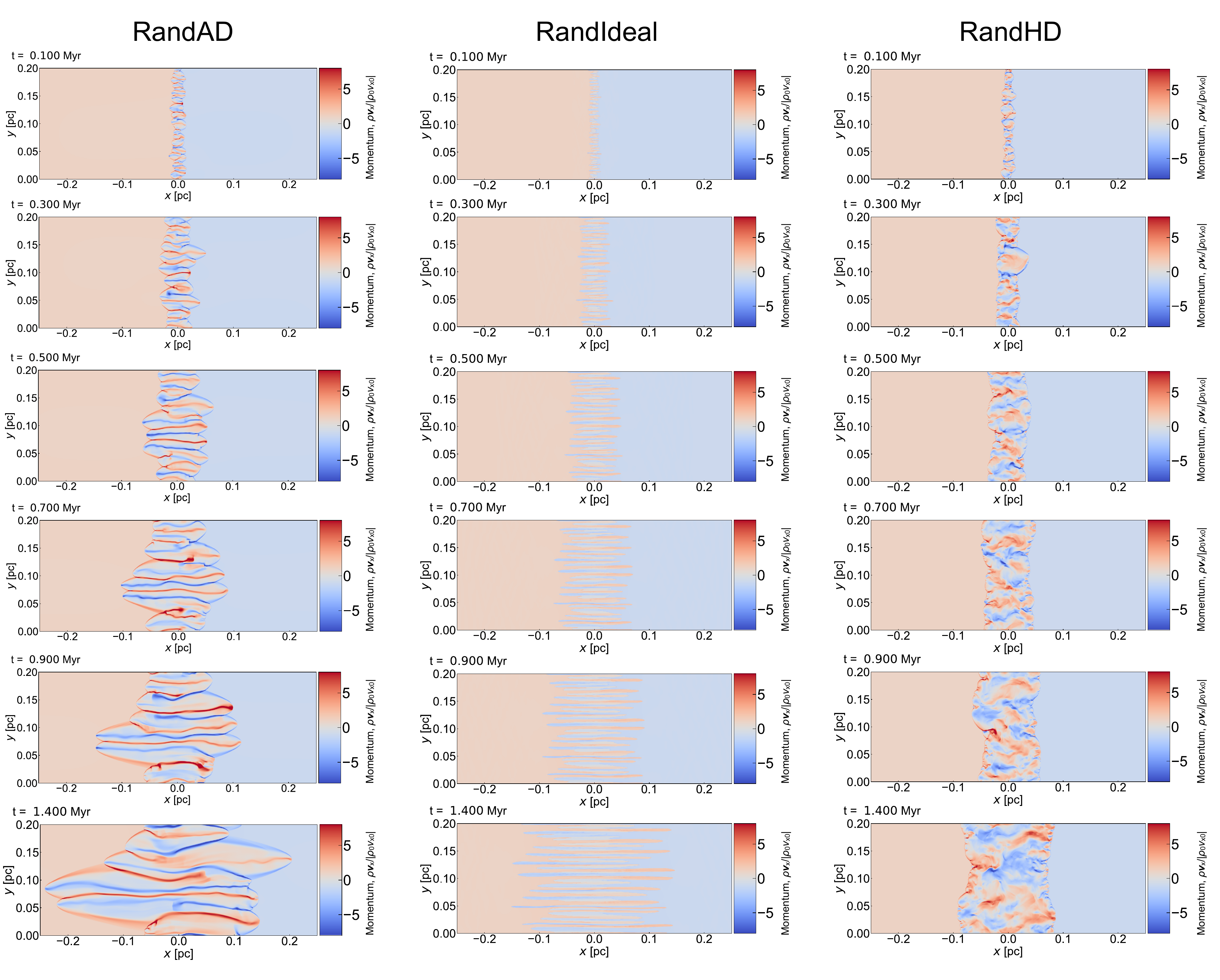}
\caption{\small{$x$-componet of momentum maps produced by models RandAD, RandIdeal, and RandHD at time $t$ = 0.1, 0.3, 0.5, 0.7, 0.9, and 1.4 Myr (top to bottom). Due to the STORM, large-scale shear motions appear even when random perturbation is used (left panels). In the ideal MHD case, the momentum transportation stops at the shock fronts (center panels). A more complex and incoherent flow structure appears in pure hydro case (right panels), which is much different compared to the results of RandAD.\label{fig:momxRand}
}}
\end{figure*}

\begin{figure*}[]
\epsscale{1.15}
\plotone{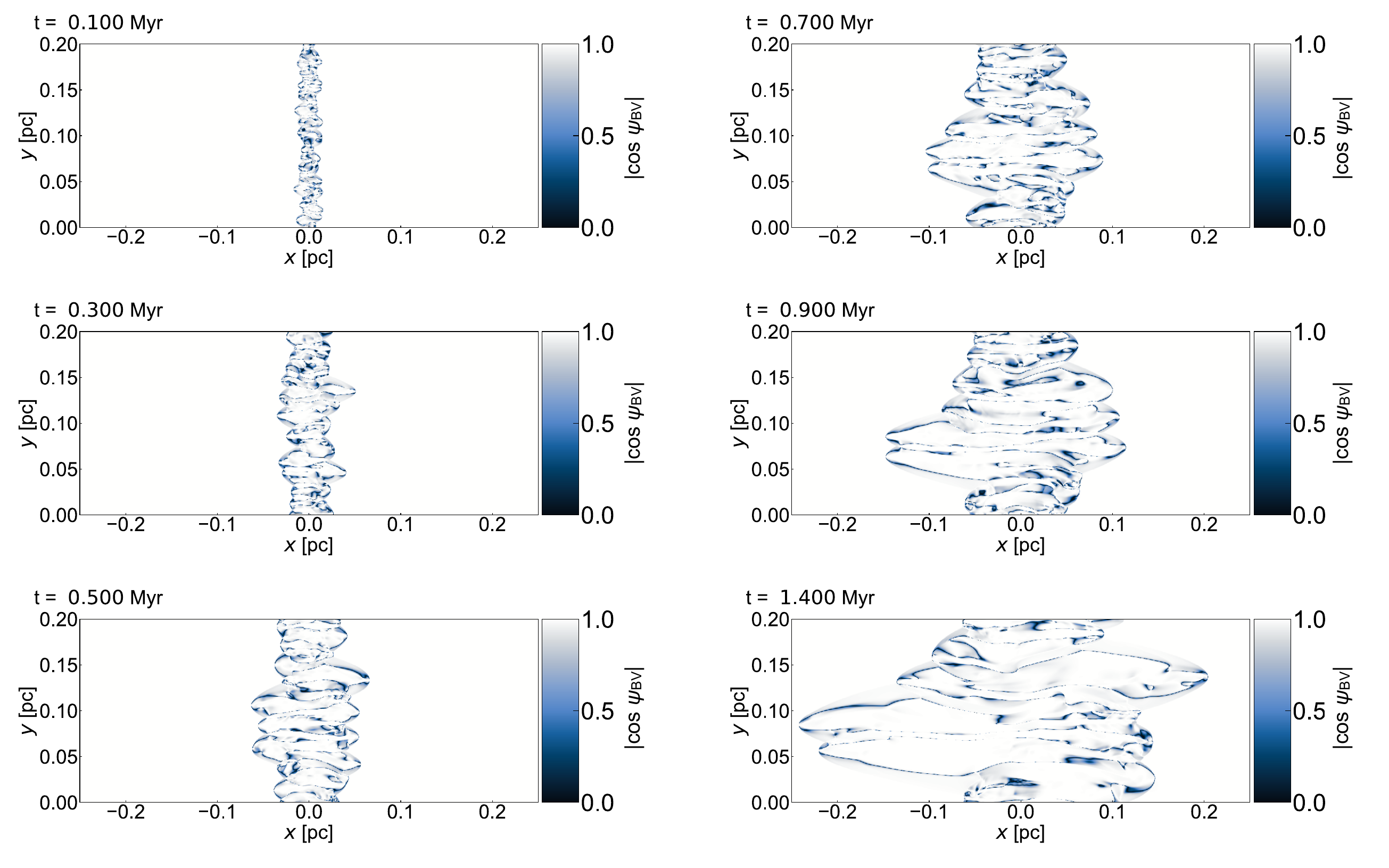}
\caption{
\small{
{$|\cos \Psi_{\mathrm{BV}}|$ maps produced by model RandAD at time $t$ = 0.1, 0.3, 0.5 (left), 0.7, 0.9, and 1.4 Myr (right). The {initial} direction of the magnetic field is parallel (or anti-parallel) in the initial velocity field.
Ambipolar diffusion is effective in the vicinity of the shock fronts and the hails (blue region) even when random shock corrugation is applied.
}
}\label{fig:2dcosPsiBVrand}
}
\end{figure*}

\subsection{Setup for the Two-dimensional Simulations} \label{subsec:setup2d}

We numerically solve Eqs. (\ref{eqs:eoc})--(\ref{eqs:ie}) on a 2D domain of size [-0.8 pc, 0.8 pc] $\times$ [0 pc, 0.2 pc].
We first focus on the nonlinear growth of SSI, ignoring self-gravity.
In this scenario, the initial density and pressure field are uniform; that is, $\rho(x,y) = \rho_0$ and $p(x,y) = p_0$.
We set the initial $x$ component of the velocity field as
\begin{equation}
v_x(x, y)=-v_{x 0} \tanh \left[\frac{x-\xi \sin \left(k_{\rm y} y\right)}{0.003\ \mathrm{pc}}\right]
\end{equation}
where, $\rho_0, v_{x0},$ $p_0$, $k_{\rm y} = 2\pi / 0.05 pc$ and $\xi$ = 0.02 pc denote the initial density, $x$-component of the velocity, pressure in the pre-shock region, the wave number corresponding to the unstable mode for SSI including ambipolar diffusion, and the corrugation amplitude, respectively.
{Here we choose the corrugation amplitude ($\xi$ = 0.02 pc) smaller than 0.1 pc.
The SSI is expected to be stable for fluctuations smaller than 0.033 pc under the parameters of the SingleModeAD model, as demonstrated in \citet{Abe2024ApJ...961..100A}. Therefore, we set the corrugation wavelength to 0.05 pc.}
In more realistic cases, the initial velocity field is given by
\begin{equation}
v_x(x, y)=-v_{x 0} \tanh \left[\frac{x- \xi \sum_{l=1}^{64}\sin \left(k_{\mathrm{y}} y + \phi_l \right)}{0.003\ \mathrm{pc}}\right],
\end{equation}
where, $k_{\mathrm{y},l}\equiv 2\pi l/ L_{\mathrm{box,y}}$, $\phi_{l}$, and $\xi$ are the wave number, a random phase, and the corrugation amplitude with a white-noise spectrum, respectively.
We select small $\xi$ = 0.0003 pc {to observe the parameter dependence clearly}, giving a collision surface dispersion of 0.0024 pc, and set the upstream gas sound speed $c_{\mathrm{s}}$ to 0.2 km s$^{-1}$ in $p_0=\rho_0 c_{\mathrm{s}}^2 / \gamma$.
Under these initial conditions, a filament forms at around $x$ = $0$.
As star-forming filaments are perpendicular to the magnetic field, the initial uniform magnetic field is set along the $x$-axis ($B_0\hat{x}$).
Figure \ref{fig:inicon2d} shows an example of the initial conditions for the 2D simulations.
{We run a set of 13 simulations by varying the parameter values of the initial conditions.
The initial conditions of the different runs are listed in Table \ref{tab:Model parameters to study the nonlinear evolution.}.}
In these models named ``Rand," the initial shock wavefront fluctuations have a white-noise spectrum.
The RandHD model is a non-magnetized reference model.

The numerical domain is a 2D box with a uniform grid of 4096 $\times$ 512 cells, giving a spatial resolution of $\Delta x$~=~$\Delta y$~=~3.9~$\times$~10$^{-4}$~pc.
{The cell size is sufficiently small to resolve the most unstable scale of SSI including the ambipolar diffusion ($\sim \ell_{\rm AD}$).}
Zero-gradient boundary conditions (assuming continuous gas flow) are imposed at the boundaries $x = -0.8$ pc and $x = 0.8$ pc and periodic boundary conditions at $y$ = 0, 0.2 pc.

\subsection{Setup for the Three-dimensional Simulations} \label{subsec:setup3d}
Previous studies of filament formation suggest that massive filaments are created in a shock-compressed layer~\citep{Pineda2023ASPC..534..233P}.
Magnetically directed inflows in the shock-compressed layer create and feed the filament~\citep[][]{inoue2013ApJ...774L..31I, Inoue2018PASJ...70S..53I, abe2021ApJ...916...83A}.
Figure \ref{fig:inicon} is a schematic of this situation.
To realistically capture the nonlinear evolution of a filament in 3D simulations, we consider self-gravity in a cubic numerical domain with side lengths of $L_{\mathrm{box,x}}$= 1.6 pc, $L_{\mathrm{box,y}}$ = 0.8 pc, and $L_{\mathrm{box,z}}$ = 1.6 pc.
As the initial condition, we set a dense sheet (Figure \ref{fig:inicon}) formulated as
\begin{equation}
n(x, y, z)=\left(n_{\rm {sheet }}-n_{\rm {ext }}\right) \exp \left(-\frac{z^2}{2 H^2}\right)+n_{\rm {ext }},
\end{equation}
where, $n_{\rm sheet}$ = 4000 cm$^{-3}$, $n_{\rm ext}$ = 5 cm$^{-3}$ are the number densities of the dense sheet and ambient gas, respectively, and $H$ = 0.1 pc is the sheet thickness.
{Here, we do not form the sheet by the first shock compression but study the formation of the filament through converging flows within the compressed sheet.}
To set converging gas flows with a corrugated colliding surface, we set the $x$ component of the velocity field as
\begin{equation}
\begin{aligned}
v_x(x, y, z)
&= -v_{x 0} \tanh \Bigg( \\ 
&\frac{x - \xi \sum_{l=1}^{24} \sum_{m=1}^{3} 
\sin \left(k_{\mathrm{y}} y + \phi_l \right) \sin \left(k_{\mathrm{z}} z + \phi_{m} \right)}{0.0006\ \mathrm{pc}} \Bigg), \\
&\quad \text{if } -H / 2 \leq z \leq H / 2.
\end{aligned}
\end{equation}
where, $k_{\mathrm z}\equiv 2\pi m/ H$ and $\phi_{m}$ are the wave number and a random phase, respectively.
$v_{x 0}$ is set to be 1.2 km s$^{-1}$.
Here, We choose $\xi$ = 0.001 pc, giving a collision surface dispersion of 0.004~pc, and set the velocity fields of the $y$ and $z$ components as $v_y$($x$, $y$, $z$) =  $v_z$($x$, $y$, $z$) = 0 km s$^{-1}$.
The initial pressure field is $p(x, y, z) = \gamma \rho(x, y, z)c^2_{\rm s}$.
We set the initial uniform magnetic field along the $x$-axis and its value is $B_0$~=~50~$\rm{\mu}$G, {which is consistent with observed magnitude considering the density in the sheet~\citep[e.g.,][]{Crutcher2012, Heiles2005a}.}
The left panel of Figure \ref{fig:inicon} is a schematic of the initial condition.
{The simulation continues until 1.16 Myr when local collapse occurs.}

We adopted a static mesh refinement technique to resolve the created filament while reducing the computational costs.
{At least the region defined by \(|x| < 0.3 \, \mathrm{pc}\) and \(|z| < 0.05 \, \mathrm{pc}\) is refined twice from the base grid (the base grid resolution is of \(\simeq 0.013 \, \mathrm{pc}\)).
The finest resolution ($x$ $\simeq$ 1.6$\times$10$^{-3}$ pc) bottlenecks in solving ambipolar diffusion term because the time step $\Delta t$ in a parabolic equation is proportional to $\Delta x^2$.}
To alleviate this bottleneck, we employ the super time stepping method~\citep{Meyer2014JCoPh.257..594M}, setting the maximum time step ratio parameter as 1000 as justified by \cite{Abe2024ApJ...961..100A}.

\section{Results} \label{sec:Results}

\subsection{Two-dimensional simulations without self-gravity} \label{subsec: Two-dimensional simulations w/o self-gravity}

The left and right panels of Figure \ref{fig:2ddensity005} show the density maps generated by the SingleModeIdeal (left column) and SingleModeAD (right column) models, respectively, at time $t$ = 0.1, 0.2, 0.3, and 0.4 Myr (top to bottom).
In both models, the SSI causes the growth of the corrugation.
{Note that, as shown in \citet{Abe2024ApJ...961..100A}, the SSI occurs when the wavelength is 0.05 pc even if there is the effect of the ambipolar diffusion. (The value of ``0.09 pc" in eqs. \ref{eqs:ambi scale} is a rough estimation.) Moreover, the SSI is expected to be stable with fluctuations smaller than 0.033 pc in this case.}
The expanding speed of the shocked layer is more prominent in model SingleModeAD than in the ideal MHD model SingleModeIdeal.
The reason for this can be explained as follows.
Ambipolar diffusion permits flow along the shock fronts and across the magnetic fields only near the shock fronts (right panel in Figure \ref{fig:twoShock}; see also the right panel of Figure \ref{fig:singleShock} for the single-shock case), forming dense blobs in the valleys known as \textit{shock channel}~\citep{SnowHillier2021MNRAS.506.1334S}.
{To demonstrate that the ambipolar diffusion induces flows in the vicinity of the shock fronts, we show $|\cos \Psi_{\mathrm{BV}}(x,y)| \equiv \left| \boldsymbol{B}(x,y) \cdot \boldsymbol{v}(x,y) / B(x,y)v(x,y) \right|$ maps, where, $\Psi_{\mathrm{BV}}$ is the angle between magnetic field and velocity field.
These maps are produced by model SingleModeAD (left panels in Figure \ref{fig:momx}).
Since the magnetic field direction is initially parallel (or anti-parallel) to the velocity, the effect of ambipolar diffusion allows for flow across the magnetic field, {leading to small values of the $|\cos \Psi_{\mathrm{BV}}|$.}
Although the ambipolar diffusion is effective throughout the entire shocked layer at $t$ = 0.1 Myr since the layer is initially thin, it does not affect the center $x \sim 0$ pc at $t$ = 0.4 Myr.
Thus, we confirm that {the} ambipolar diffusion mainly impacts the vicinity of the shock fronts.}
The blobs flow toward the center of the shocked layer and stir the gas inside the filament, injecting momentum (see {right panels} in Figure \ref{fig:momx}).
Momentum transport causes internal turbulence and faster expansion of the compression layer than for the ideal MHD model.
{Note that the blobs are already present on the convex parts of the wavefront and exert pressure on the wavefronts. The formation and ejection of blobs begin at the very start of the simulation. The prominent ejection of blobs seen at $t$ = 0.2, 0.3, and 0.4 Myr are secondary events, created within the small concave regions at the leading edge of the shock front.}
Here, we name this {turbulence-driving} mechanism the {\textit{STORM} (Slow-shock-mediated Turbulent flOw Reinforced by Magnetic diffusion) and the dense blob the \textit{hail} because the dense blobs resemble hails that are ejected like a hailstorm into the filament} (see Figure \ref{fig:mechssiad} for a schematic illustration).
The STORM transforms the kinetic energy of the converging flows into anisotropic turbulence in the filament, providing effective pressure to increase the filament width.
If the effect of ambipolar diffusion is neglected, i.e., the STORM does not occur {(because the gas cannot move across the magnetic field lines)}, the kinetic energy of the converging flows is converted into thermal energy, which is immediately removed via radiative cooling.
Figure \ref{fig:twoShock} shows the detailed velocity structures in the SingleModeIdeal and SingleModeAD models.
Most of the converging flow gas stops at the shock fronts in model SingleModeIdeal but the gas moves through the post-shock region in SingleModeAD.

\begin{figure*}[ht!]
\epsscale{0.9}
\plotone{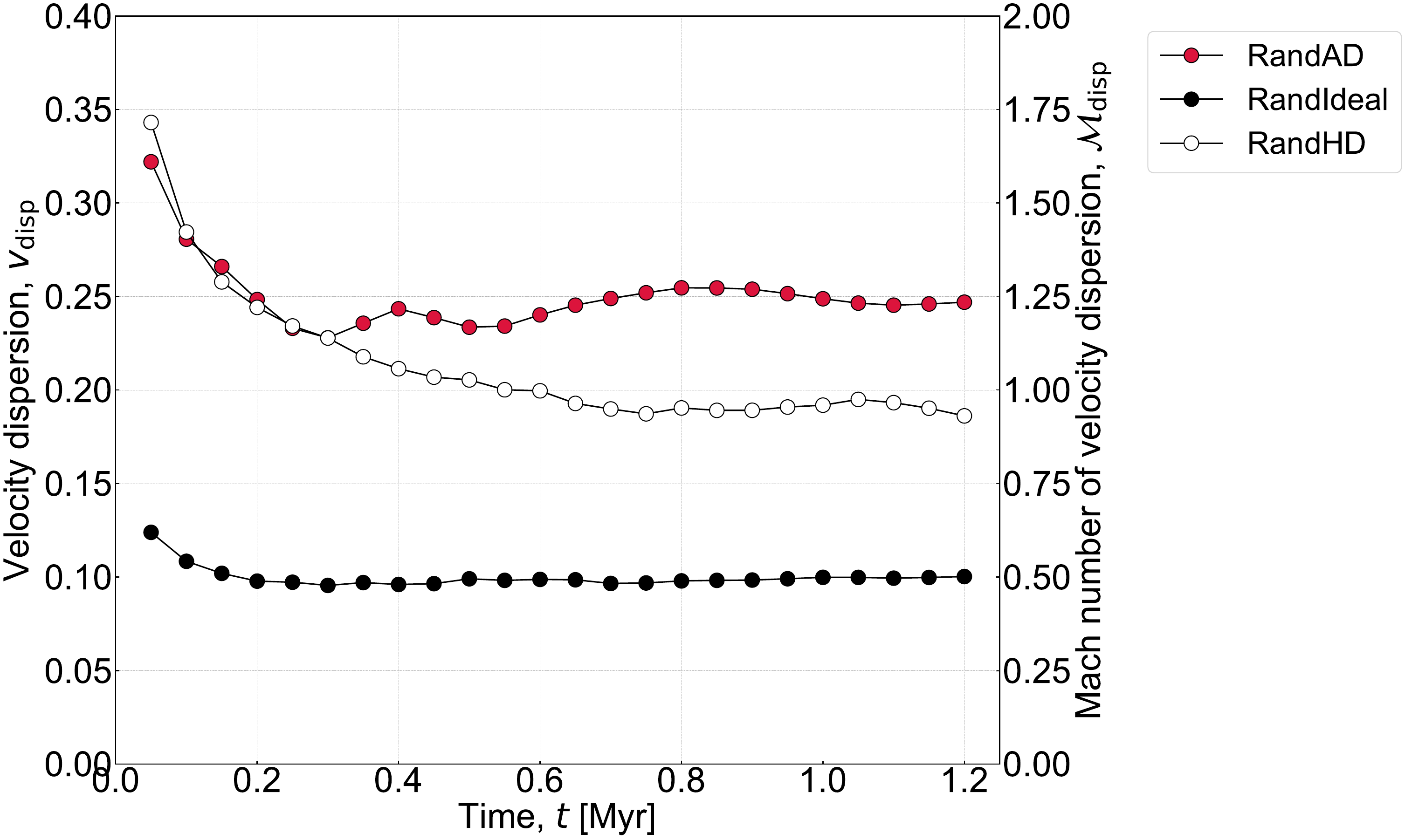}
\caption{\small{Evolutions of the turbulent Mach number in the post-shock region, i.e., within the filament, given by models RandIdeal (ideal MHD model, black circles), RandAD (non-ideal MHD model, red circles), and RandHD (unmagnetized model, white circles) without the effect of self-gravity. 
In the RandIdeal model, the shocks are unstable but the strict frozen-in condition inhibits the turbulence of the gas motion.
In the RandHD model, nonlinear thin-shell instability induces turbulence in the post-shock region at the initial stage (white circles), but as the shocked slab thickens,
instability growth is halted, and the turbulence decays. In the RandAD model, the momentum transportation continues and the corrugation of the shock fronts grows because the SSI remains unstable (red circles).
\label{fig:vdisp}
}}
\end{figure*}

More realistic cases with random initial shock corrugations are simulated in RandIdeal, RandAD, and RandHD.
The results are displayed in Figure \ref{fig:2ddensityrand}.
Owing to the {STORM} (ambipolar diffusion effect), the post-shock region expands faster in RandAD (left panel of Figure \ref{fig:2ddensityrand}) than in the case without ambipolar diffusion (RandIdeal, center panel) and pure hydro case (RandHD, right panel).
{We also show {the} $x$-componet of momentum maps produced by models RandAD, RandIdeal, and RandHD. Due to the STORM, large-scale ordered motions appear even when random shock corrugation is used (left panels).
In the ideal MHD case, the momentum transportation stops at the shock fronts (center panels).
Owing to {the} nonlinear thin-shell instability~\citep{Vishniac1994ApJ...428..186V},
a more complex and incoherent flow structure appears in the pure hydro case (right panels), which is significantly different compared to the results of RandAD.
To support the claim that ambipolar diffusion impacts the vicinity of the shock fronts, even when random shock corrugation occurs, we show $|\cos \Psi_{\mathrm{BV}}|$ maps produced by models RandAD (Figure \ref{fig:2dcosPsiBVrand}).
Flows guided by the magnetic fields (white region) can be observed at the inside of the shocked layer, especially at $t$ = 1.4 Myr.
We can confirm that ambipolar diffusion impacts the vicinity of the shock fronts (blue region) even when random shock corrugation is used.}
The velocity dispersion $v_{\mathrm{disp}}$ at $t$ = 1.0 Myr in the compression layer (defined as the high-density region with $>$ 5$\times \rho_0$) {is larger by factors $\sim$ 2 -- 20} in RandAD ($v_{\mathrm{disp}}$ $\simeq$ 0.18 km s$^{-1}$) than in RandIdeal and RandHD ($v_{\mathrm{disp}}$ $\simeq$ 0.0078 km s$^{-1}$ and 0.095 km s$^{-1}$, respectively).
These values do not notably change over time (see Figure \ref{fig:vdisp}).
In the RandIdeal model, the shocks are unstable but the turbulence of the gas motion is inhibited by the strict frozen-in condition.
The velocity dispersion remains relatively small under the saturation effect of SSI in ideal MHD~\citep{StoneEdelman1995ApJ...454..182S}.
In the RandHD model, nonlinear thin-shell instability induces turbulence in the post-shock region at the initial stage  (white circles in Figure \ref{fig:vdisp}), but it is known that instability growth is halted and the turbulence decays as the shocked slab thickens\citep{Vishniac1994ApJ...428..186V}.
In the RandAD model, the momentum transportation continues and the corrugation of the shock fronts grows because the SSI remains unstable (red circles in Figure \ref{fig:vdisp}).
Note that the compressed layer continues to expand in all 2D cases because the self-gravity is not included.
{To quantify the contribution of hails, we have examined the ratio of the kinetic energy of the hails to the total kinetic energy in model RandAD within the filament. While this result depends on the threshold density $\rho_{\rm threshold}$ used to define the hails, the ratio is 0.56 (0.22) for $\rho_{\rm threshold}$ = 10$\rho_0$ (15$\rho_0$). This indicates that both the turbulent pressure generated by the hails and the shear motion within the filament contribute to the dynamics.}
In conclusion, the persistence and efficiency of turbulence inside the filament are driven by the magnetic field and the ambipolar diffusion effect, respectively.

\subsubsection{Parameter Dependence}
\begin{figure*}[ht!]
\epsscale{1.2}
\plotone{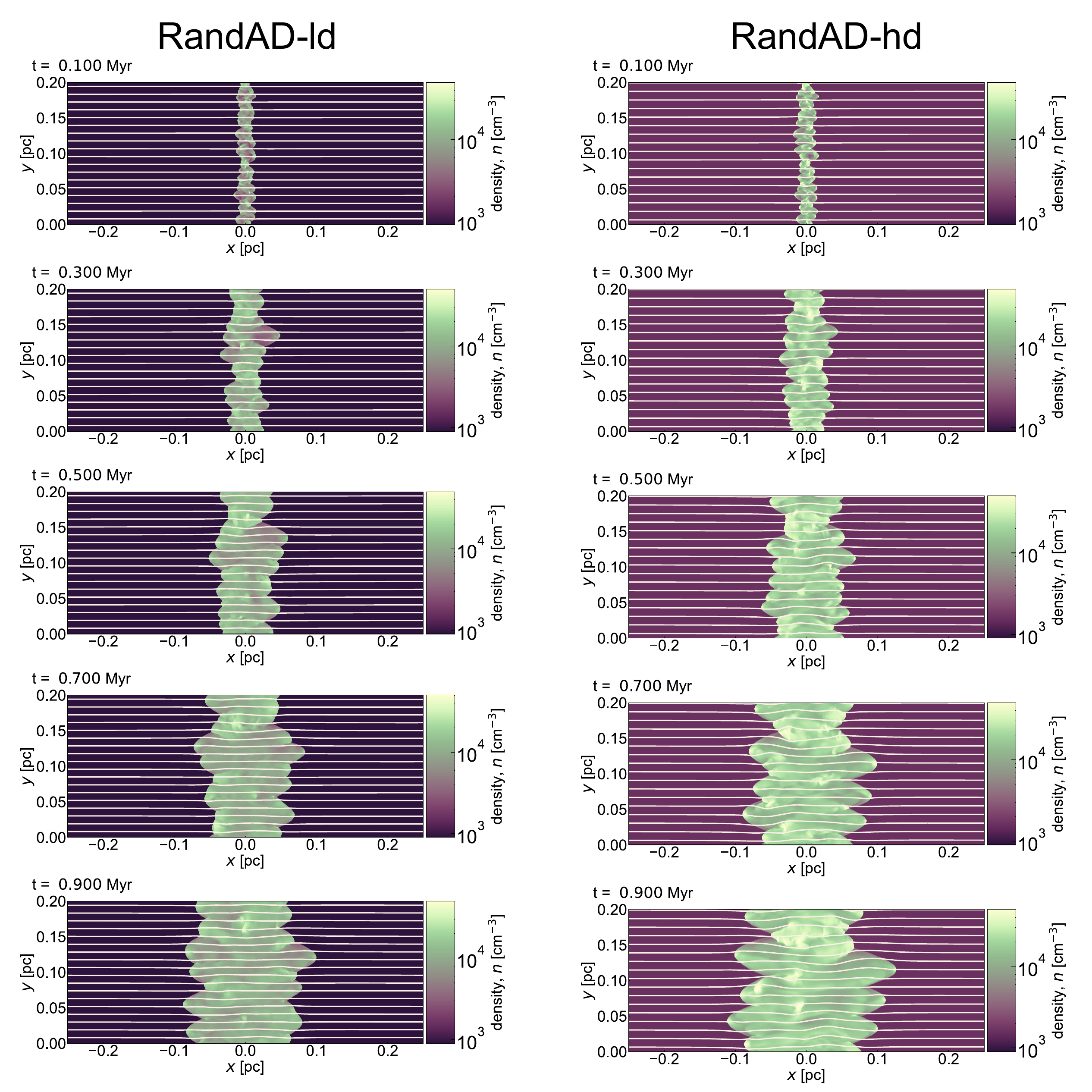}
\caption{\small{Density maps generated by models RandAD-ld ($n_0=800$ cm$^{-3}$, left column) and RandAD-hd ($n_0=1600$ cm$^{-3}$, right column) with different initial densities at time $t$ = 0.1, 0.3, 0.5, 0.7, and 0.9 Myr (top to bottom). White lines represent the magnetic field lines.
{The expansion speeds remain similar when we change the density.}
\label{fig:2ddensity_n}
}}
\end{figure*}
\begin{figure*}[ht!]
\epsscale{1.2}
\plotone{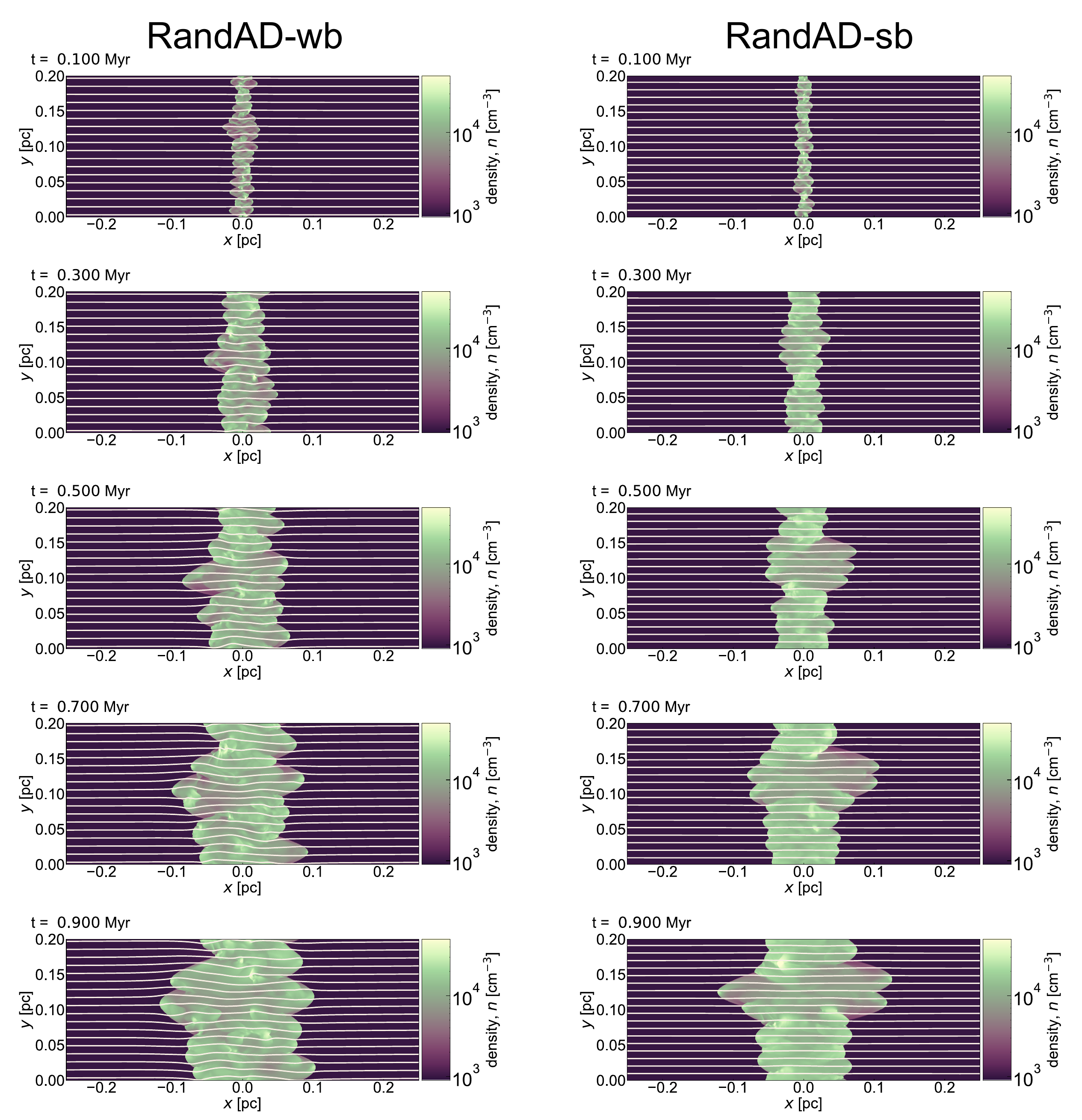}
\caption{\small{Similar to Figure \ref{fig:2ddensity_n} but for models RandAD-wb ($B_0 = 24$ $\rm{\mu}$G, left column) and RandAD-sb ($B_0 = 50$ $\rm{\mu}$G, right column) with different magnetic field strengths.
{The expansion speeds remain similar when we change the strength of the magnetic field.}
\label{fig:2ddensity_b}
}}
\end{figure*}
\begin{figure*}[ht!]
\epsscale{1.2}
\plotone{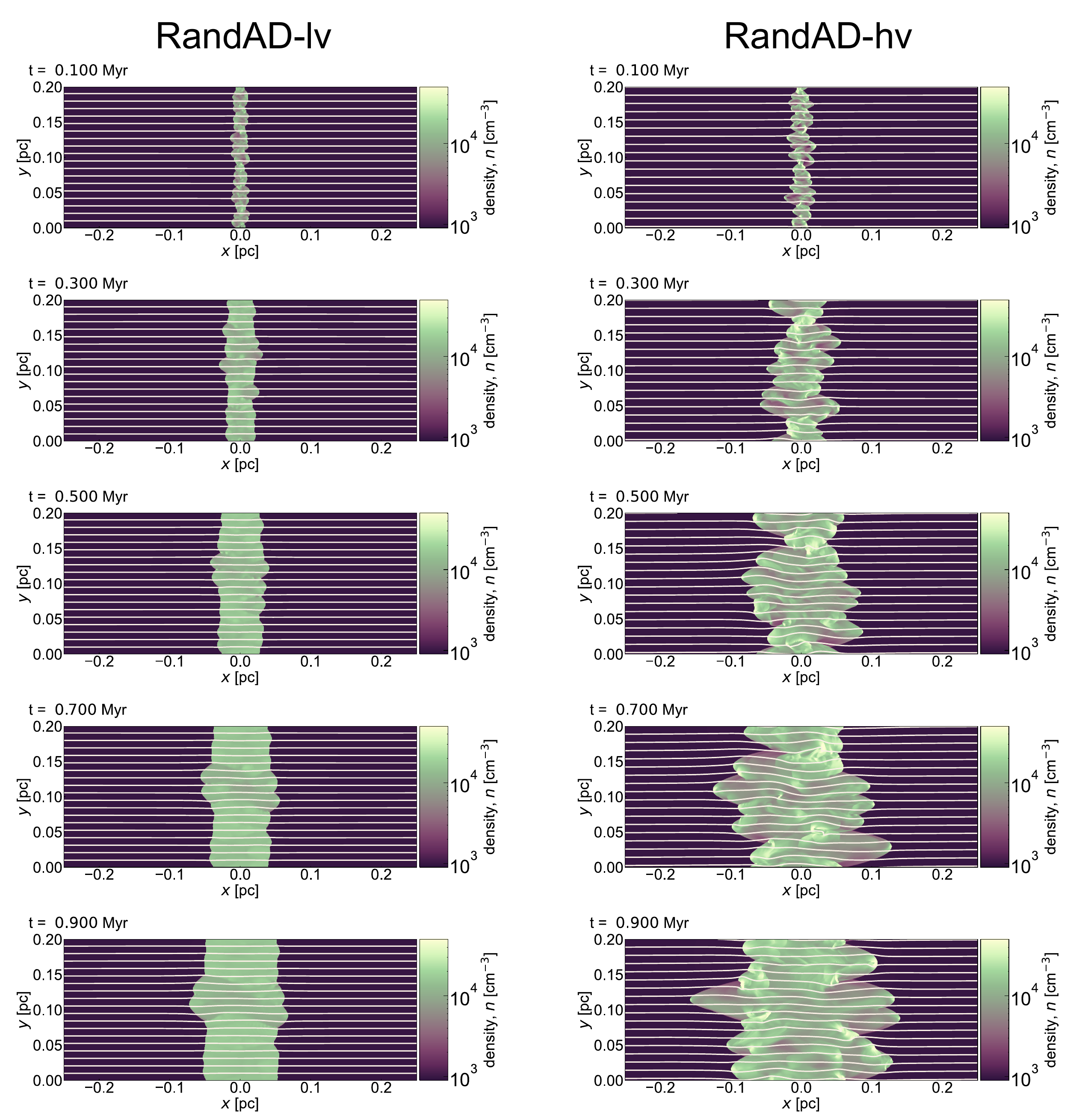}
\caption{\small{Similar to Figure \ref{fig:2ddensity_n} but for models RandAD-lv ($v_0=0.8$ km s$^{-1}$, left column) and RandAD-hv ($v_0=1.3$ km s$^{-1}$, right column) with different initial velocities.
{Increasing the converging flow velocity increases the expansion speed of the shocked region.}
\label{fig:2ddensity_v}
}}
\end{figure*}
\begin{figure*}[ht!]
\epsscale{0.9}
\plotone{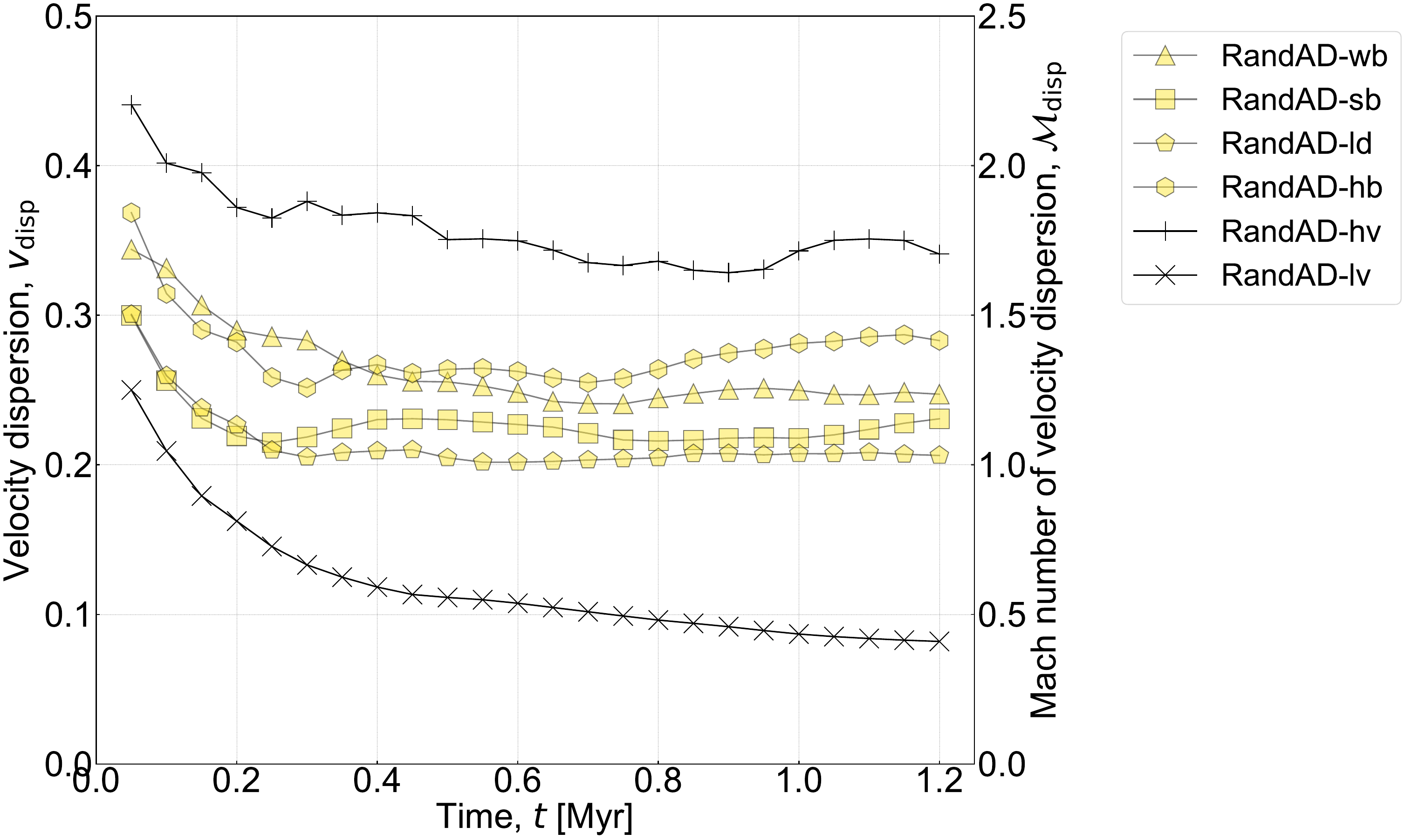}
\caption{\small{Evolution of the turbulent Mach number in the post-shock regions of models RandAD-ld ($n_0=800$ cm$^{-3}$, yellow pentagons), RandAD-hd ($n_0=1600$ cm$^{-3}$, yellow hexagons), RandAD-wb ($B_0=24$ $\rm{\mu}$G, yellow triangles), RandAD-sb ($B_0=50$ $\rm{\mu}$G, yellow squares), RandAD-lv ($v_0=0.8$ km s$^{-1}$, black pluses), and RandAD-hd ($v_0=1.3$ km s$^{-1}$, black crosses).
{Although the density and magnetic field strength minimally affect the conversion rate of inflow velocity to turbulent velocity dispersion (yellow symbols), we observe that the inflow velocity controls the velocity dispersion (black symbols). }
\label{fig:vdisp_pram}
}}
\end{figure*}
\begin{figure*}[ht!]
\epsscale{1.17}
\plotone{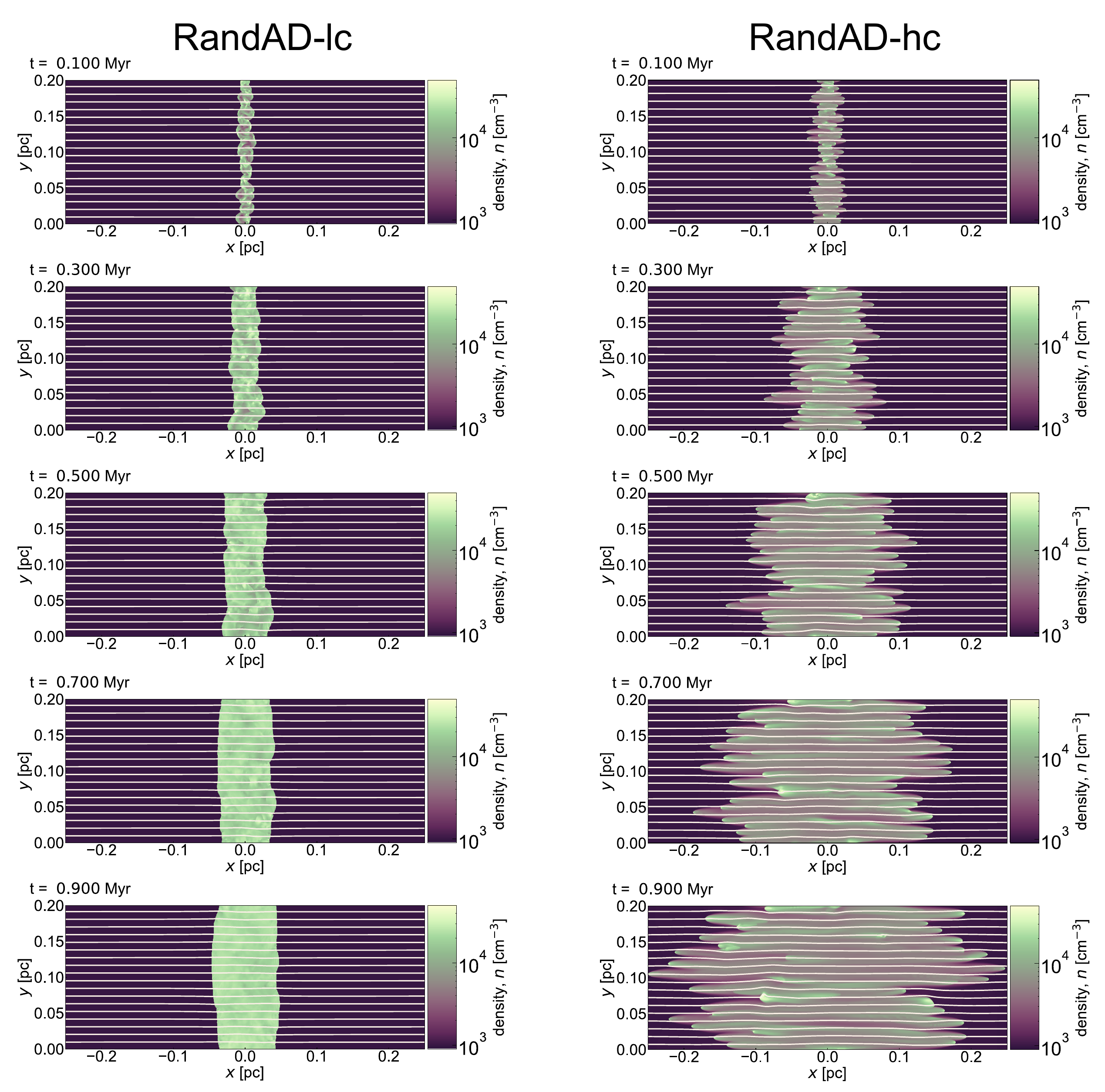}
\caption{\small{
{
Similar to Figure \ref{fig:2ddensity_n} but for models RandAD-lc (left panels, $C=3\times 10^{-17}$ cm$^{-3/2}$g$^{1/2}$) and RandAD-hc (right panels, $C=3\times 10^{-15}$ cm$^{-3/2}$g$^{1/2}$) with different ambipolar diffusion coefficients.
A lower ionization degree suppresses the STORM mechanism (left panels), while a higher ionization degree produces finer finger structures and smaller hail sizes (right panels).
\label{fig:2ddensity_eta-lh}
}
}}
\end{figure*}
\begin{figure*}[ht!]
\epsscale{1.17}
\plotone{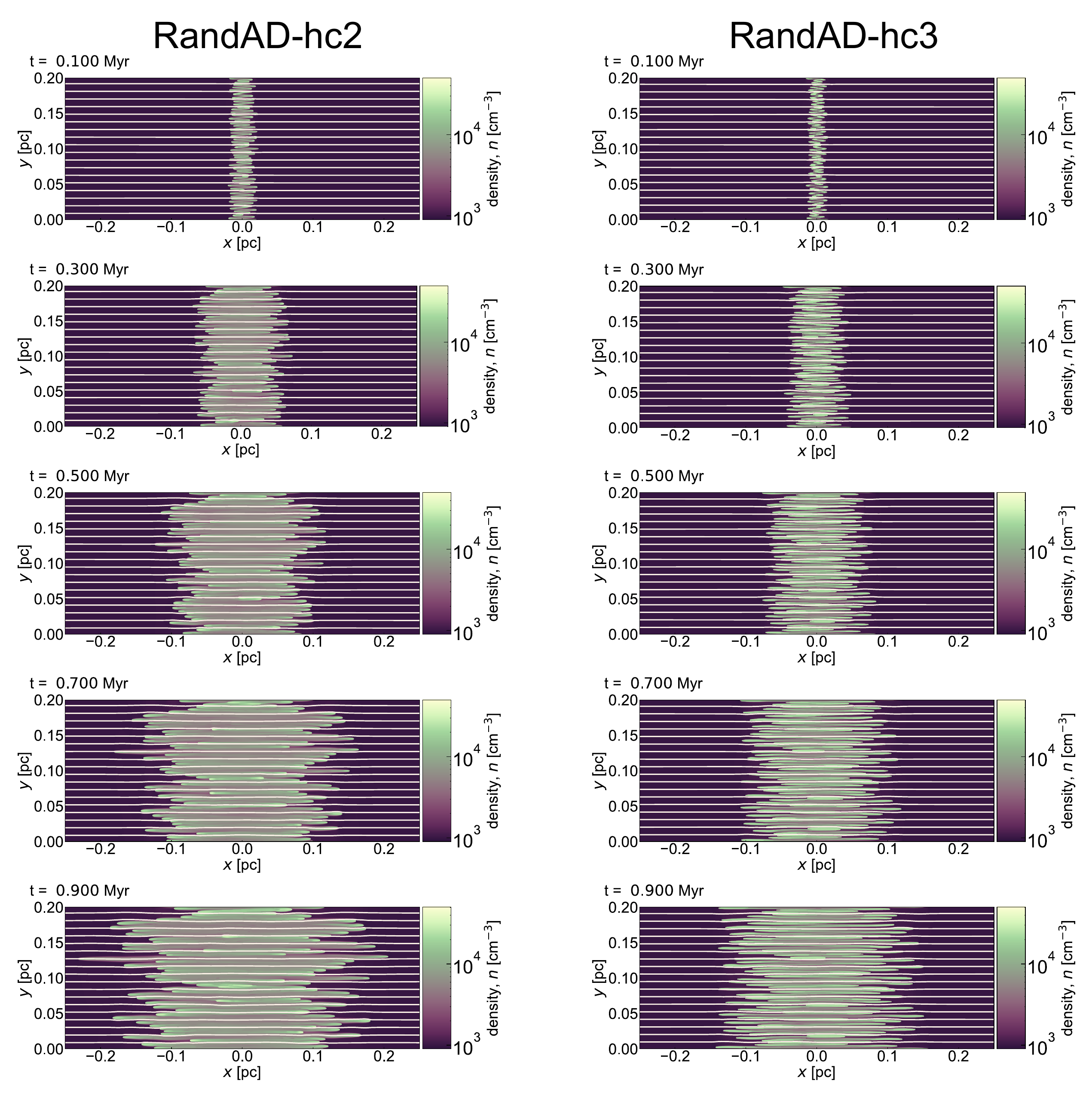}
\caption{\small{
{
Similar to Figure \ref{fig:2ddensity_n} but for models RandAD-hc2 ($C=3\times 10^{-14}$ cm$^{-3/2}$g$^{1/2}$) and RandAD-hc3 ($C=3\times 10^{-13}$ cm$^{-3/2}$g$^{1/2}$) {with much lower ambipolar diffusion coefficients than fiducial ionization degree model of RandAD.
The lower ionization degree produces a finer structure compared to model RandAD-hc, and the STORM mechanism remains effective.}
\label{fig:2ddensity_eta-h23}
}
}}
\end{figure*}
\begin{figure*}[ht!]
\epsscale{0.9}
\plotone{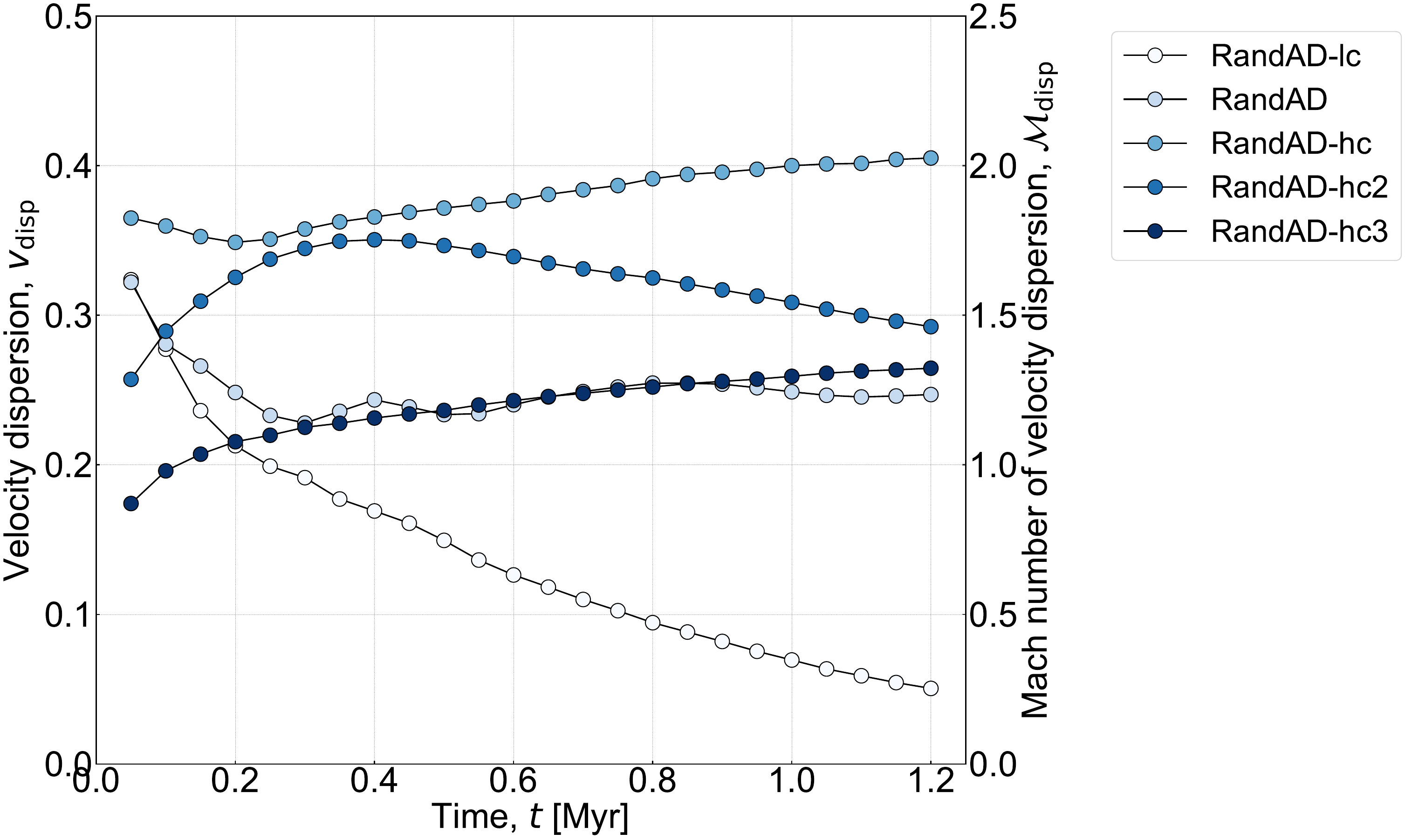}
\caption{\small{Similar to Figure \ref{fig:vdisp_pram} but for models RandAD-lc (white, $C=3\times 10^{-16}$ cm$^{-3/2}$g$^{1/2}$), RandAD (light blue), RandAD-hc (blue, $C=3\times 10^{-15}$ cm$^{-3/2}$g$^{1/2}$), and RandAD-hc2 (dark blue), $C=3\times 10^{-14}$ cm$^{-3/2}$g$^{1/2}$. Other parameters are the following. $n_0 = 1000 \,\mathrm{cm^{-3}}, B_0 = 30\,\mathrm{\mu G}, v_{\mathrm{x0}} = 1.0\,\mathrm{km s^{-1}}$.
{The velocity dispersion within the filament remains unchanged within a factor of 2 across a wide ionization degree range, particularly in high-ionization degree models. 
For the low-ionization degree model (RandAD-lc), the STORM is not effective.}
\label{fig:vdisp_eta}
}}
\end{figure*}

To study the dependence of the important parameters, we vary the density, magnetic field, shock velocity, and ambipolar diffusion coefficient (ionization degree) from the fiducial model.
Figure \ref{fig:2ddensity_n} shows {the density distribution and the magnetic field lines} at t = 0.10 0.30, 0.50, 0.70, and 0.90 Myr in models RandAD-ld and RandAD-hd with different densities.
Equivalent results for models RandAD-wb and RandAD-sb are shown in Figure \ref{fig:2ddensity_b}.
The expansion speeds remain quite similar when we change the density and magnetic field strength.
For a more quantitative discussion, Figure \ref{fig:vdisp_pram} shows the evolution of velocity dispersion in the shocked layer in models RandAD-ld, RandAD-hd, RandAD-wb, and RandAD-sb (yellow symbols).
The density and magnetic field strength minimally affect the conversion rate of inflow velocity to turbulent velocity dispersion (which remains around 25\%), but obviously, affect the length scale of ambipolar diffusion (see Eq. \ref{eqs:ambi scale}) and hence the {hail} size and mass because the length scale of ambipolar diffusion corresponds to the thickness of the gas accumulation region in the shock channel.
Therefore, although higher density gas or weaker magnetic field strength refine the structure and {reduce the spacing of the hails} in the post-shock region, the velocity dispersion in the filaments remains similar.

Figure \ref{fig:2ddensity_v} presents snapshots of the density map in models RandAD-lv and RandAD-hv at t = 0.10 0.30, 0.50, 0.70, and 0.90 Myr.
Increasing the converging flow velocity increases the expansion speed of the shocked region.
From the velocity dispersion in the shocked layer (black symbols in Figure \ref{fig:vdisp_pram}), where we observe that the {inflow} velocity controls the velocity dispersion.
This provides an insight into the velocity dispersion of turbulence, {which is roughly given by $\Delta v =f v_x$, where, $f \sim 0.25$ is the conversion factor from the {inflow} velocity to the generated internal turbulence.}
Note that $f$ is a constant given by the simulation result, and independent of density and magnetic field.
In the low-velocity case {(0.8 km s$^{-1}$), the} ambipolar diffusion has a significant impact, resulting in a smoother structure within the shock-compressed region.
Increasing the upstream velocity increases the magnetic Reynolds number hence the fineness of the structures in the post-shock region {(in other words, the spacing between the hails reduces)}.


{To investigate the influence of the ionization degree, we created models RandAD-lc, RandAD-hc, RandAD-hc2, and RandAD-hc3 with $n_0$ = $1000$ cm$^{-3}$ and ionization degrees of $\sim$~3.9$\times$10$^{-8}$ ($C=3\times 10^{-17}$ cm$^{-3/2}$ g$^{1/2}$), 3.9$\times$10$^{-6}$ ($C=3\times 10^{-15}$ cm$^{-3/2}$ g$^{1/2}$), 3.9$\times$10$^{-5}$ ($C=3\times 10^{-14}$ cm$^{-3/2}$ g$^{1/2}$), and 3.9$\times$10$^{-4}$ ($C=3\times 10^{-13}$ cm$^{-3/2}$ g$^{1/2}$) respectively.}
The density maps and time-dependent velocity dispersion results are displayed in Figures \ref{fig:2ddensity_eta-lh}, \ref{fig:2ddensity_eta-h23} and \ref{fig:vdisp_eta}, respectively.
{The low-ionization model (RandAD-lc) shows a reduced velocity dispersion due to the suppression of the STORM mechanism. Interestingly, the velocity dispersion is even lower than that in the pure hydro model (RandHD).
This behavior can be attributed to the influence of the magnetic field, which helps stabilize the nonlinear thin-shell instability~\citep[see also][which investigates the non-linear thin-shell instability in a magnetized medium]{Heitsch2007ApJ...665..445H}.}
The STORM is effective in high ionization degree models (RandAD-hc, RandAD-hc2, and RandAD-hc3), as evidenced by the high-velocity dispersion in Figure \ref{fig:vdisp_eta}.
The RandAD-hc, RandAD-hc2, and RandAD-hc3 models differ from the fiducial model RandAD only in the scale of their hails and corrugations of shocks (right panels of Figure \ref{fig:2ddensity_eta-lh} and Figure \ref{fig:2ddensity_eta-h23}) as expected from Eq. (\ref{eqs:ambi scale}).
{Note that considering that the velocity dispersion of the extremely high-ionization degree model (RandAD-hc3) declines compared to model RandAD-hc and approaches the velocity dispersion in the model (RandAD), a result of an even higher ionization model will approach the ideal MHD.}
{On the other hand, the ambipolar diffusion coefficient does not affect the total momentum injected into the filament, because the increase or decrease of the hail size is compensated by the reduction or the enhancement of the total number and {the spacing} of hails within the filament.
Consequently, the velocity dispersion within the filament is not notably affected {by a broad range of the ionization degree} (see Figure \ref{fig:vdisp_eta})}.


\subsection{Three-dimensional simulations with self-gravity} \label{subsec: Three-dimensional simulations w/ self-gravity}
\begin{figure*}[ht!]
\epsscale{1.17}
\plotone{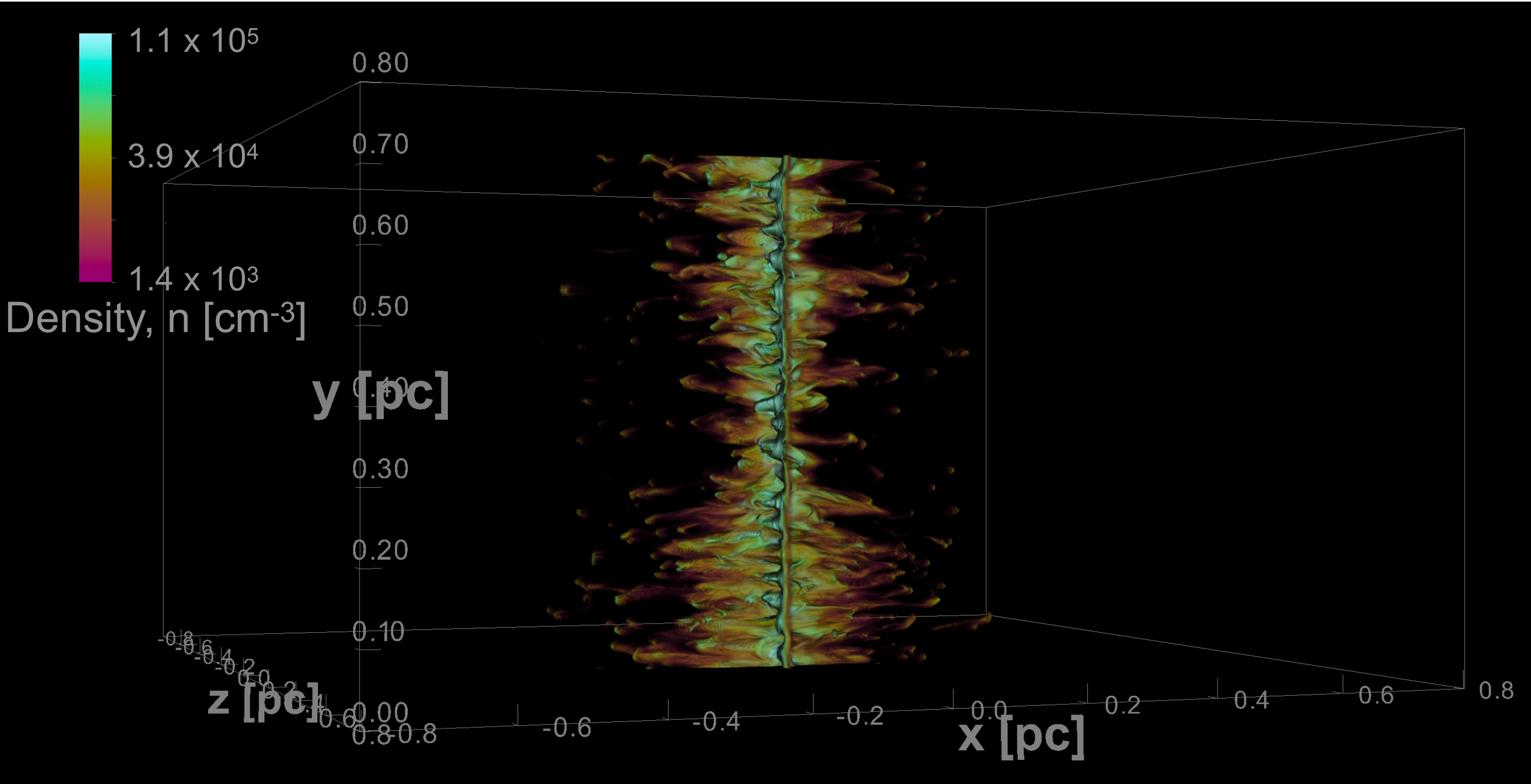}
\caption{\small{Volume rendering plot of density of the 3D SSI simulation with self-gravity and ambipolar diffusion at $t$ = 1.15 Myr\label{fig:3dfil}
}}
\end{figure*}
\begin{figure*}[ht!]
\epsscale{1.1}
\plotone{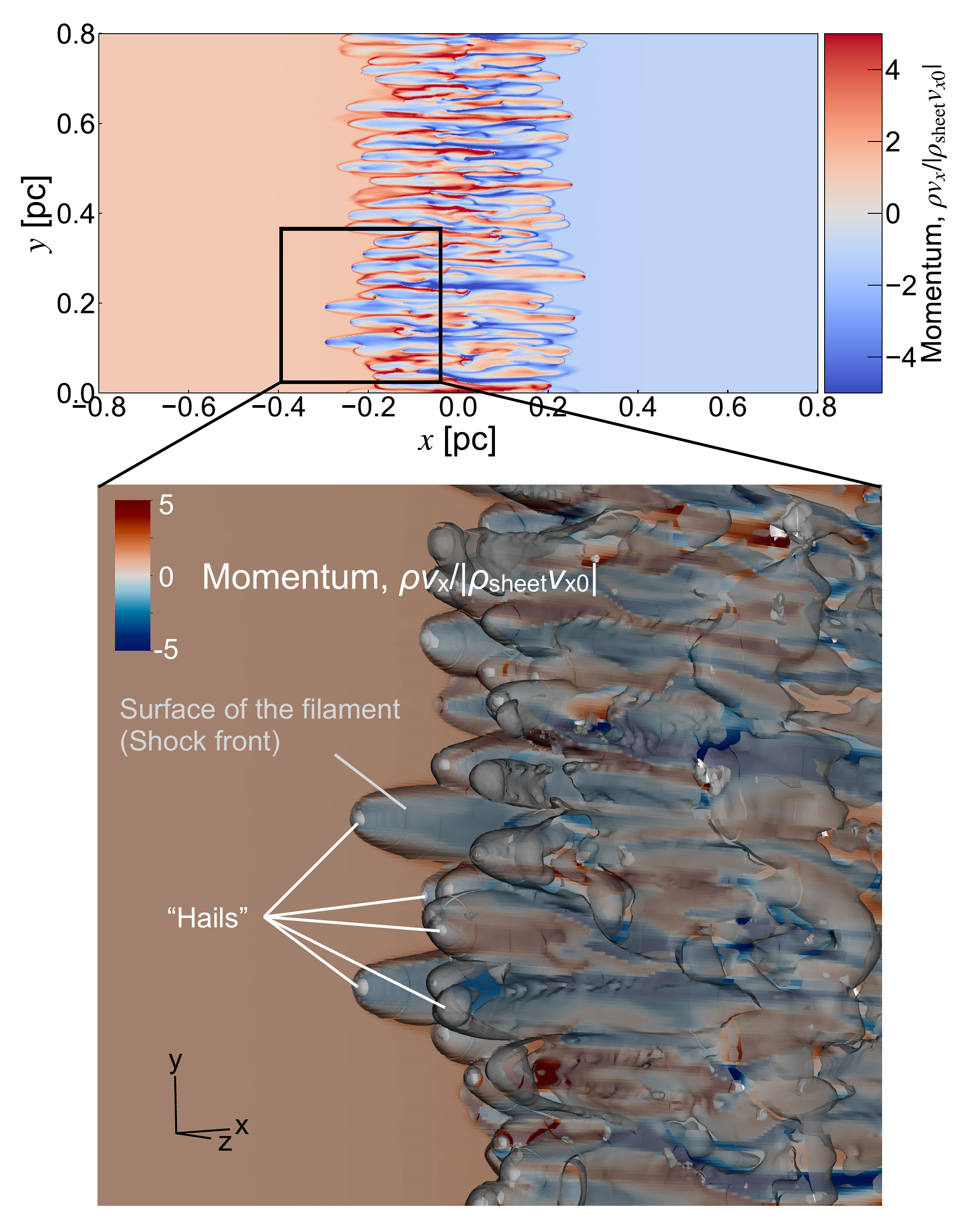}
\caption{\small{
{
\textit{Top}: The $x$-component momentum slice at $z$ = 0 pc from the 3D SSI simulation with self-gravity and ambipolar diffusion at $t$ = 0.55 Myr.
\textit{Bottom}: A similar slice plot to the top panel, along with 3D contour plots of density at 6.8$\times 10^3$ cm$^{-3}$ (gray, approximately corresponding to the surface of the filament) and 6.8$\times 10^4$ cm$^{-3}$ (white, corresponding to hails).
The STORM feature is clearly observed even in the 3D simulation.
}
\label{fig:3dmomx}
}}
\end{figure*}
\begin{figure*}[ht!]
\epsscale{1.15}
\plotone{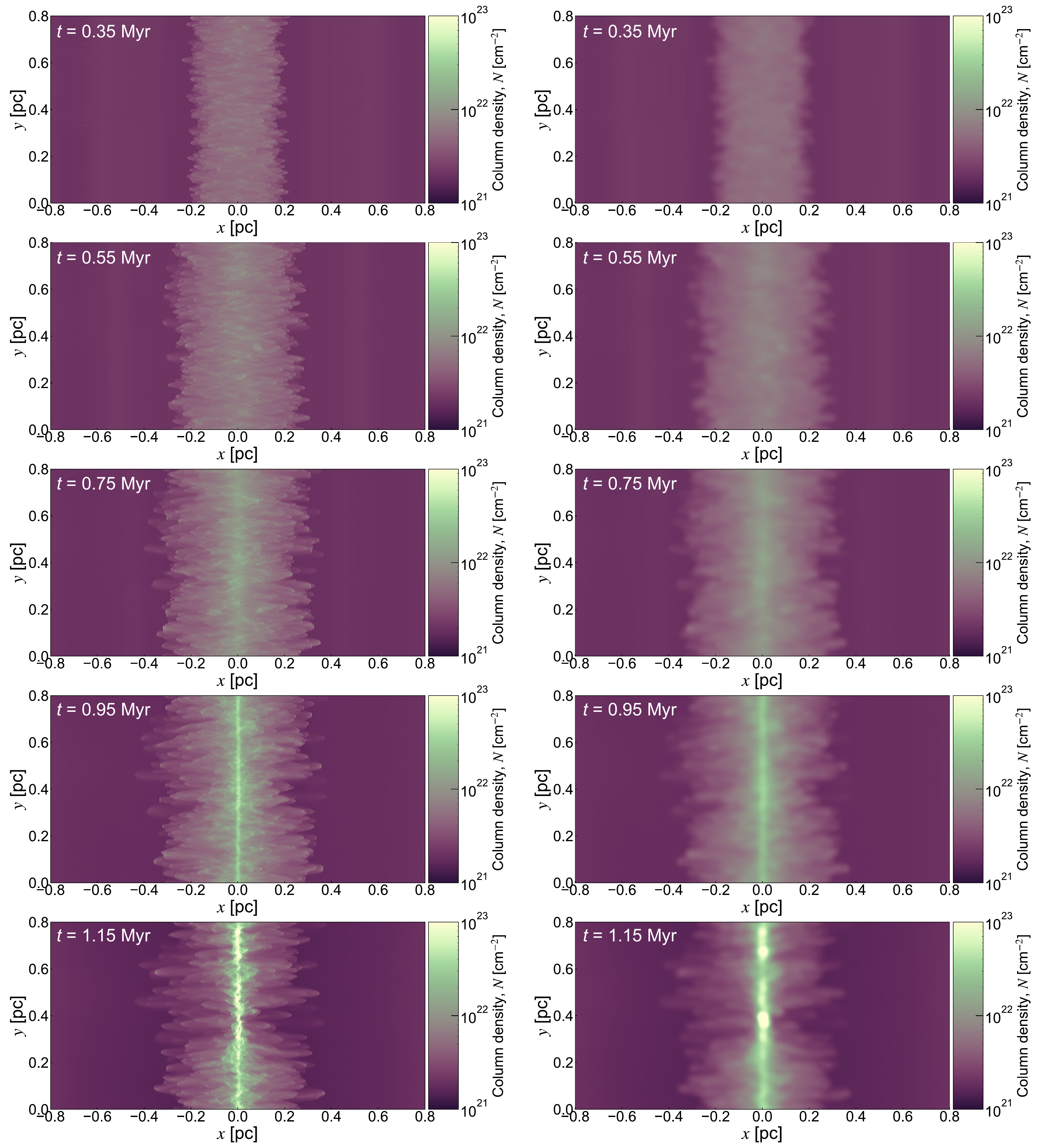}
\caption{\small{\textit{Left}: Column density map obtained in the 3D SSI simulation with self-gravity and ambipolar diffusion at time $t$ = {0.35, 0.55, 0.75, 0.95, and 1.15 Myr}. The resolution of the finest level is 0.0016 pc. Wavefront corrugation and hails similar to those in the 2D simulations are observed.
\textit{Right}: The same column density map after smoothing with a Gaussian kernel function of width {0.012 pc}, to be compared with the filaments observed with \textit{Herschel} in nearby regions at $\sim$ 140 pc distances~\citep{Arzoumanian2019A&A...621A..42A}.
\label{fig:column3d}
}}
\end{figure*}

\begin{figure*}[ht!]
\epsscale{1.17}
\plotone{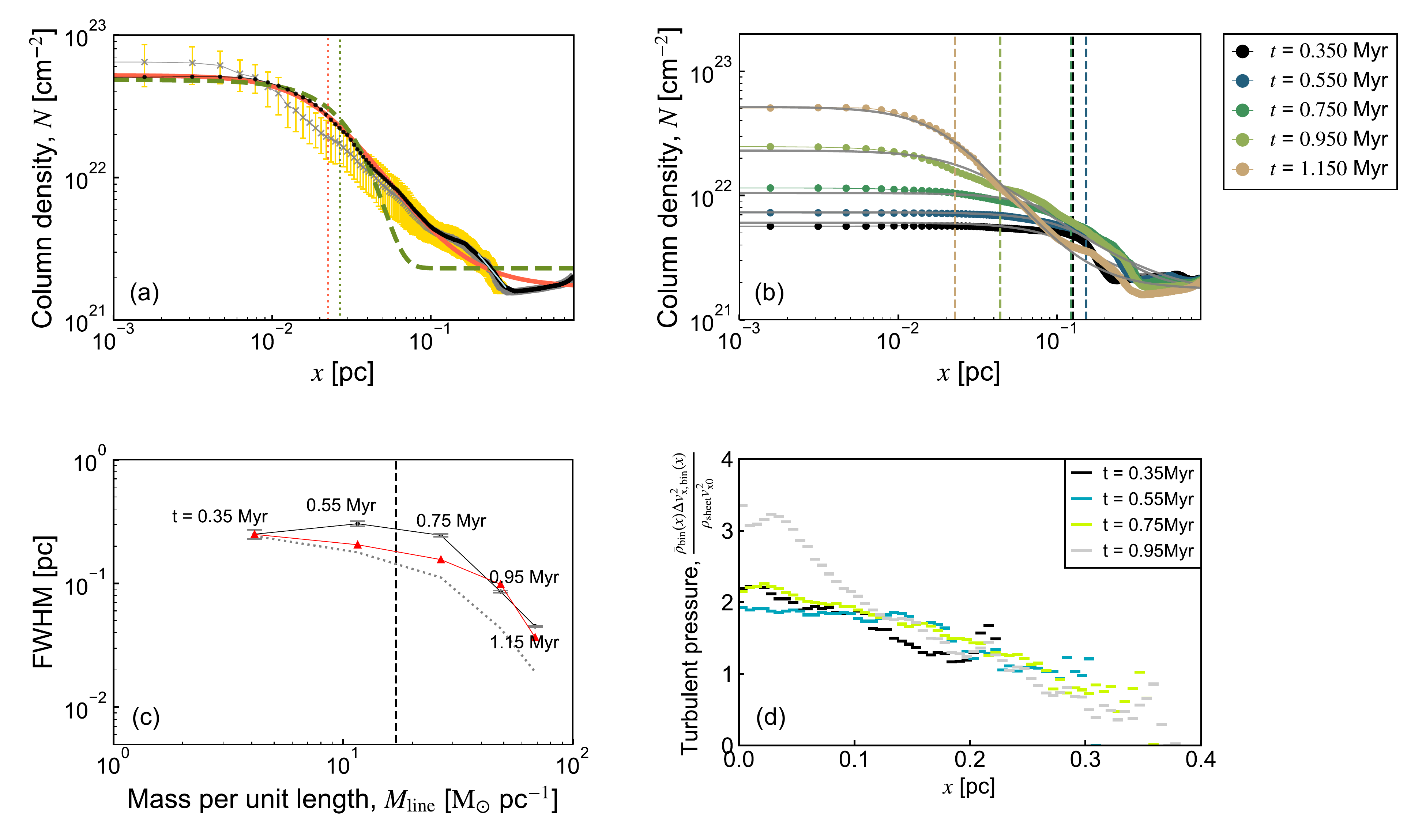}
\caption{\small{{
{
(a) Radial column density profile across the filament ($t$ = 1.15 Myr) in the x-direction parallel to the magnetic field averaged along the y-axis (gray crosses), the median absolute deviation of the distributions of independent cuts taken perpendicular to the filament crest (yellow error bars), the smoothed column density profile using a 0.012 pc wide Gaussian kernel function (black dots), Plummer function (red line), half the full-width half maximum derived from the Plummer function fitting (FWHM/2, i.e., the radius of the filament) (red dotted line), Gaussian function (green dash line), half the FWHM from the Gaussian function fitting (green dotted line), and the fitting curve in the range 0.2 pc $<$ $x$ $<$ 0.3 pc (light-blue line).
The steepening of the column density profiles in the 0.2 pc $<$ x $<$ 0.3 pc region can be attributed to the slow-shock.
(b) {Results for multiple simulation epochs (the time is indicated on the panel).} Smoothed radial column density profile across the filament in the x-direction parallel to the magnetic field averaged along the y-axis (colored dots), Plummer function (gray lines), FWHM/2 (colored dashed lines).
(c) FWHM-line mass diagram. The vertical dashed line represents the thermal critical line mass. The grey error bars represent the uncertainties in the FWHM, calculated based on the variances of $R_{\rm flat}$ and $p$.
The black dotted line shows the thermal scale height defined by $\sqrt{2c_{\rm s}^2 / (\pi G \bar{\rho}_{\rm 0})}$ [see Eq. (\ref{eqs: density radial profile for isothermal cylinder}) or Eq. (2) in \citet{InutsukaMiyama1992}] for $T$ = 10 K at $t$ = 0.35 and 0.55 Myr from left to right and the red dotted line shows the effective scale height defined by $\sqrt{2(c_{\rm s}^2 + \Delta v^2) / (\pi G \bar{\rho}_{\rm 0})}$ at $t$ = 0.35, 0.55 0.75, 0.95, and 1.15 Myr from left to right.
The filament evolves along the track from left to right. Our study shows that a thermally supercritical filament maintains its width for a certain duration.
(d) Radial turbulent pressure profiles. We measure the mean densities and velocity dispersions in each spatial bin with a width of 6.4$\times 10^{-3}$ pc and define the turbulent pressure as $\bar{\rho}_{\mathrm{bin}} \Delta v^2_{\mathrm{x,bin}}$. The turbulent pressure gradient driven by the STORM mechanism acts as a supporting force.
}
\label{fig:radialprofiles}
}}}
\end{figure*}
\begin{figure}[ht!]
\epsscale{1.1}
\plotone{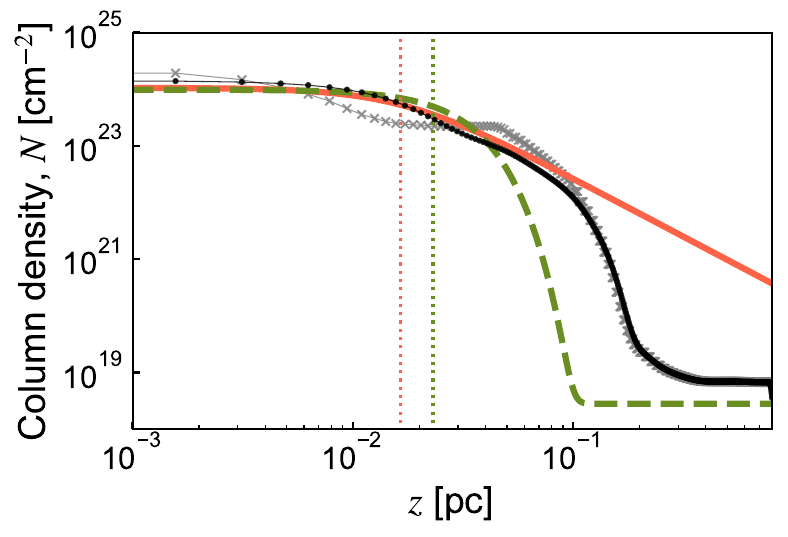}
\caption{\small{{
{
Similar to the panel (a) in Figure \ref{fig:radialprofiles} for the filament radial profile in the $z$-direction perpendicular to the sheet.
The FWHM value perpendicular to the sheet given by the Plummer function fitting $w_{\rm FWHM, \perp}$ is 0.033 pc, which is close to the FWHM measured along the magnetic field direction (0.045 pc).
Our study shows that a supercritical filament with a line mass larger by a factor of four compared to the thermal critical value formed by the STORM has a width of $\sim$ 0.1 pc.
}
\label{fig:radialprofiles_z}
}}}
\end{figure}

Figure \ref{fig:3dfil} shows the density plot generated in the 3D MHD simulation with self-gravity and ambipolar diffusion at $t$ = 1.15 Myr when the filament sufficiently gains mass.
We measure the line mass of the filament by tracing the dense region of $n$ $>$ 5$n_{\mathrm{sheet}}$ = 20000 cm$^{-3}$, and obtain $M_{\rm line}$ = 68 $M_{\odot}$ pc$^{-1}$, exceeding by a factor four the thermal critical value of $\sim 17 M_{\odot}$ pc$^{-1}$ for 10 K (see a detailed explanation and a discussion below).
{To demonstrate that a wavefront corrugation and hails similar to those observed in the 2D simulations appear even in the 3D simulation, we show the $x$-component momentum map at $z$ = 0 pc (top panel of Figure \ref{fig:3dmomx}) and the 3D density contour plot (bottom panel of Figure \ref{fig:3dmomx}).
We select $t$ = 0.55 Myr to capture the state before the self-gravity strongly sets, which allows a direct comparison with the 2D results.
The wavefront corrugation and hail features, resembling those in the 2D simulations, clearly emerge.
The velocity dispersion in the dense region remains nearly constant over time, with a value of 0.62 km s$^{-1}$, leading to a conversion factor $f \sim 0.5$, where, the velocity dispersion is defined as $\Delta v \equiv \sqrt{\Delta v_x^2 + \Delta v_y^2 + \Delta v_z^2}$, with $\Delta v_x$, $\Delta v_y$, and $\Delta v_z$ representing the velocity dispersions in the $x$, $y$, and $z$ directions, respectively.}
The reasons for the increased conversion efficiency (between the 2D and 3D simulations) may be attributed to the difference in the spatial dimension and its detail will be investigated in a future work.
The left panels of Figure \ref{fig:column3d} show the column density map obtained by integrating the density field along the $z$-direction.
{Interestingly, \citet{Tachihara2024ApJ...968..131T} have reported that fine ``feather"-like substructures within the filament, which is similar to the substructure seen in the left panels in Figure \ref{fig:column3d}, are observed in a high-density region of the Corona Australis cloud.
They compared the observation and our simulation result and concluded that our simulation reproduces the observed ``feather" structures.}


{In order to compare with the observed width of filaments, we create the right panels in Figure \ref{fig:column3d} showing the result of smoothing the column density maps using a 0.012 pc wide Gaussian kernel function, which corresponds to the physical resolution of the \textit{Herschel} images in the nearby clouds at a distance of $\sim$ 140 pc~\citep{Arzoumanian2019A&A...621A..42A}.
Panel (a) in Figure \ref{fig:radialprofiles} shows the column density profile (gray crosses) and its smoothed profile (black dots).}
The radial column density profile averaged along the $y$-axis is given by:
\begin{equation}
N_{\rm profile}(x)= {\rm median}_{y=0\ \rm{pc}}^{L_{\mathrm{box,y}}}N(x,y),\ 
\mathrm{for}\ x \geq 0.0\ \rm{pc},
\end{equation}
where, $N(x,y)$ is the column density along the $z$-axis, and the ${\rm median}_{y=a}^{b}$ operator takes a median value in the range $a<y<b$.
{The column density profile is constructed in the finest grid scale.}
{The local dispersion of the radial $N(x,y)$ profiles (yellow error bars) is estimated as the median absolute deviation.}
Following \citet{Arzoumanian2011A&A...529L...6A, Arzoumanian2019A&A...621A..42A}, we fit the profile with the following Plummer function and one-dimensional Gaussian function:
\begin{equation}
N_{\mathrm{Pl}}(r)= \frac{N_{\mathrm{0,Pl}}}{\left(1+\left(r / R_{\rm {flat}}\right)^2\right)^{(p-1) / 2}} + N_{\mathrm{bg,Pl}},
\end{equation}
\begin{equation}
N_{\mathrm{G}}(r)= N_{\mathrm{0,G}} \exp \left[-\frac{r^2}{2 \sigma^2}\right] + N_{\mathrm{bg,G}},
\end{equation}
where, $N_{\mathrm{0,Pl}}$, $R_{\rm flat}$, $p$, $N_{\mathrm{bg,Pl}}$, $N_{\mathrm{0,G}}$, $\sigma$, and $N_{\mathrm{bg,G}}$ are free fitting parameters.
{Applying least squares fitting to $N_{\rm profile}(x)$ derived from our simulation smoothed to a resolution of 0.012 pc (Figure \ref{fig:column3d} right), we obtain $N_{\mathrm{0,Pl}}$~=~5.1$\times 10^{22}$~cm$^{-2}$, $R_{\rm flat}$~=~0.021~pc, $p$~=~2.8, $N_{\mathrm{bg,Pl}}$~=~1.7$\times 10^{21}$~cm$^{-2}$, $N_{\mathrm{0,G}}$~=~4.6$\times 10^{22}$~cm$^{-2}$, $\sigma$~=~0.023~pc, and $N_{\mathrm{bg,G}}$~=~2.3$\times 10^{21}$~cm$^{-2}$ at $t$ = 1.15 Myr.
The free parameters are bounded as follows: $0.5 N_{\mathrm{profile}}(x=0\ \mathrm{pc}) < N_{\mathrm{0,Pl}} < 1.5 N_{\mathrm{profile}}(x=0\ \mathrm{pc})$, $0.0\ \mathrm{pc}<R_{\rm flat}<0.8\ \mathrm{pc}$, $1.2<p<4.5$, $0.5 N_{\mathrm{profile}}(x=0.8\ \mathrm{pc}) < N_{\mathrm{bg,Pl}} < 1.5 N_{\mathrm{profile}}(x=0.8\ \mathrm{pc})$, $0.5 N_{\mathrm{profile}}(x=0\ \mathrm{pc}) < N_{\mathrm{0,G}} < 1.5 N_{\mathrm{profile}}(x=0\ \mathrm{pc})$, $0.0\ \mathrm{pc}<\sigma<0.8\ \mathrm{pc}$, and $0.5 N_{\mathrm{profile}}(x=0.8\ \mathrm{pc}) < N_{\mathrm{bg,G}} < 1.5 N_{\mathrm{profile}}(x=0.8\ \mathrm{pc})$.
Our result is well described by the Plummer function but deviates at $x=0.05$ pc from the Gaussian fitting.
The full-width half-maximum (FWHM) derived from the Plummer and Gaussian function fitting are $w_{\rm FWHM,Pl}$ = {$2 R_{\mathrm{flat }}\left(2^{2 /(p-1)}-1\right)^{1 / 2}$} = 0.045 pc and $w_{\rm FWHM,G}$ = $2 \sqrt{2 \ln{2}} \sigma$ = 0.054 pc, respectively.}
Observations of typical filaments exhibit a width of $\simeq$ 0.1 pc {within a factor of 2}~\citep{Arzoumanian2011A&A...529L...6A, Arzoumanian2019A&A...621A..42A}, consistent with our result.
Our simulation setup has no background structure outside the filament since we focus on the formation of a single filament, which tends to steepen the column density profile of the filament compared to the case with a realistic background.
This could lead to a larger value of $p$ than the observed mean values of $\sim$ 2.
The steepening of the column density profiles in the 0.2 pc~$<$~$x$~$<$~0.3~pc region is attributed to the slow-shock.
Fitting the profile in Figure \ref{fig:radialprofiles} with the power law function $a_{\rm outer} x ^{p_{\rm outer}}$ in the 0.2 pc~$<$~$x$~$<$~0.3~pc region, we obtain $p_{\rm outer}=-1.3$ that {might be} used to observationally prove the slow-shock.
Note that this outer structure is more emphasized due to the idealized background, suggesting that the structure would not be easy to observe.
{In panel (b) of Figure \ref{fig:radialprofiles}, we show the time evolution of the smoothed radial profiles (colored dots), similar to those in panel (a).
The gray lines represent Plummer function fitting curves. 
The free parameters are bound in the same way as in panel (a).
The parameter $p$, obtained from the fitting, is initially very high ($t < 0.75$ Myr) because self-gravity has not yet dominated, and $p$ traces the shock jump profile.
However, at later times ($t = 0.75, 0.85, 0.95, 1.15$ Myr), after self-gravity gradually begins to take over, we find that $p$ values of 3.8, 1.8, 1.7, 2.1, and 2.8 are obtained, indicating that the value of $p$ becomes consistent with observations (1.5$<p<$2.5) in $0.75 \ \mathrm{Myr} < t < 1.15 \ \mathrm{Myr}$.
The vertical dashed lines represent the FWHM/2 derived from the Plummer fitting.
The FWHMs are replotted in panel (c) as a function of the line mass.
The grey error bars represent the uncertainties in the FWHM, calculated as $\left[\left( \partial w_{\rm FWHM,Pl} / \partial R_{\rm flat} \right)^2 \Delta R_{\rm flat}^2 + \left( \partial w_{\rm FWHM,Pl} / \partial p \right)^2 \Delta p^2 \right]^{1/2}$, where $\Delta R_{\rm flat}^2$ and $\Delta p^2$ denote the variances of $R_{\rm flat}$ and $p$, respectively, obtained from the fitting process.
The black dotted line shows the thermal scale height defined by $\sqrt{2c_{\rm s}^2 / (\pi G \bar{\rho}_{\rm 0})}$ [see Eq. (\ref{eqs: density radial profile for isothermal cylinder}) or Eq. (2) in \citet{InutsukaMiyama1992}] for $T$ = 10 K at $t$ = 0.35 and 0.55 Myr from left to right.
The red dotted line shows the effective scale height defined by $\sqrt{2(c_{\rm s}^2 + \Delta v^2) / (\pi G \bar{\rho}_{\rm 0})}$ at $t$ = 0.35, 0.55 0.75, 0.95, and 1.15 Myr from left to right. 
Where the central density $\bar{\rho}_{\rm 0}$ is defined by $\int_{y=0\ \rm{pc}} ^{L_{\rm{box,y}}} {\rho}_{\rm 0}(0,y,0) dy/L_{\rm{box,y}}$.
The filament evolves along the track from left to right.
Our study shows that the effective scale height is consistent with the FWHM, and that a thermally supercritical filament maintains its width for a certain duration.
It is worth mentioning that the supercritical filament formed by the STORM mechanism, with a line mass four times larger than the thermal critical value, retains a width of $\sim$ 0.1 pc for at least $\sim$ 1 Myr (see panel (c) in Figure \ref{fig:radialprofiles}).
The radial turbulent pressure profiles at $t$ = 0.35 (black), 0.55 (blue), 0.75 (green), and 0.95 (gray) Myr are shown in panel (d) of Figure \ref{fig:radialprofiles}.
We define the mean densities $\bar{\rho}_{\mathrm{bin}}(x)$ and velocity dispersions $\Delta v^2_{\mathrm{x,bin}}(x)$ in spatial bins for x-direction with a width of $6.4 \times 10^{-3}$ pc within the filament ($\rho > 5 \rho_{\mathrm{sheet}}$).
A pure hydro simulation performed by \citet{Heigl2020MNRAS.495..758H} showed that the accretion-driven turbulent pressure is constant across the filament, implying that the turbulent pressure does not contribute to the stability of the filaments.
However, in our simulation, the radial decline of the turbulent pressure shown in panel (d) confirms that the STORM mechanism acts as a supporting force.}

We also measured the FWHM perpendicular to the sheet by fitting the column density profile with the Plummer function and the one-dimensional Gaussian function (Figure \ref{fig:radialprofiles}, right).
The column density profile along the $z$-axis is given by:
\begin{equation}
    N_{\mathrm{profile}}(z) = N(x=0\ \mathrm{pc}, z),\ \mathrm{for}\ z \geq 0\ \mathrm{pc},
\end{equation}
where, $N(x, z)$ represents the column density given by integral along the $y$-axis, smoothed using a 0.02 pc wide Gaussian kernel.
Note that this column density profile $N_{\mathrm{profile}}(z)$ is not averaged along the $x$-direction but for $x=0$ pc slice.
{The free parameters are bound as follows: $0.5 N_{\mathrm{profile}}(z=0\ \mathrm{pc}) < N_{\mathrm{0,Pl}} < 1.5 N_{\mathrm{profile}}(z=0\ \mathrm{pc})$, $0.0\ \mathrm{pc}<R_{\rm flat}<0.8\ \mathrm{pc}$, $1.2<p<4.5$, $0.5 N_{\mathrm{profile}}(z=0.8\ \mathrm{pc}) < N_{\mathrm{bg,Pl}} < 1.5 N_{\mathrm{profile}}(z=0.8\ \mathrm{pc})$.
From the fits, we obtain $N_{\mathrm{0,Pl}} = 1.1 \times 10^{24}\ \mathrm{cm}^{-2}$, $R_{\rm flat} = 0.017\ \mathrm{pc}$, $p = 3.1$, $N_{\mathrm{bg,Pl}} = 3.2 \times 10^{18}\ \mathrm{cm}^{-2}$, $N_{\mathrm{0,G}} = 9.9 \times 10^{23}\ \mathrm{cm}^{-2}$, $\sigma = 0.020$ pc and $N_{\mathrm{bg,G}} = 2.8 \times 10^{21}\ \mathrm{cm}^{-2}$.
The Plummer function also provides a more accurate fit for high column density regions than the Gaussian function.
The Plummer function fitting deviates from our data at $z \simeq 0.1$ pc, but the deviation is small compared to that in the high column density region.
On the other hand, the Gaussian function deviates at $z \simeq 0.05$ pc. (Figure \ref{fig:radialprofiles_z}).
The values of FWHM for the Plummer and Gaussian function fits for the profile perpendicular to the sheet are $w_{\mathrm{FWHM,Pl, \perp}}$ = 0.033 pc and $w_{\mathrm{FWHM,G, \perp}}$ = 0.046 pc, which are close (or a little smaller) to the FWHM measured along the magnetic field direction ($w_{\mathrm{FWHM,Pl}}$ = 0.045 pc and $w_{\mathrm{FWHM,G}}$ = 0.054 pc).}
In conclusion, the high-line mass filament with a line mass larger by a factor of $\sim$ 4 compared to the thermal critical value formed under the influence of the STORM is highly consistent with the observed filament width.

\section{Discussion} \label{sec: Discussion}
\subsection{A Theoretical Understanding of the Origin of the Width of Supercritical Filaments}

\begin{figure}[ht!]
\epsscale{1.15}
\plotone{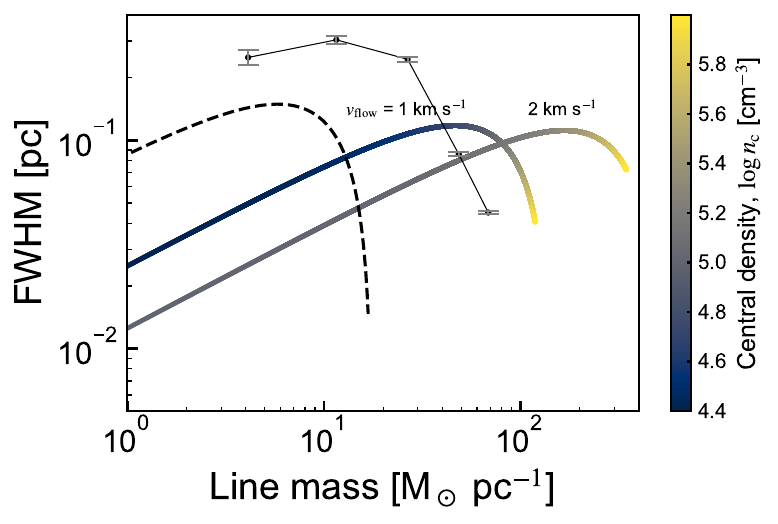}
\caption{\small{FWHM of the truncated Stod\'olkiewicz filament for $p_{\rm ext}/k_{\rm B}=2\times 10^4$ K cm$^{-3}$ as a function of line mass (dashed line).
The same curves are plotted for turbulent cases with $v_{\rm flow}=1$ and 2 km s$^{-1}$ with $n_{\rm ext}=10^3$ cm$^{-3}$.
The color scale represents the central density.
{The error-barred black dots represent the FWHM measured from the simulation as shown in Figure \ref{fig:radialprofiles} (c)}.
This model suggests that the width of the turbulent filament confined by inflow does not substantially depend on the inflow velocity, and becomes $\sim$ 0.1 pc.
\label{FilW}
}}
\end{figure}

{Based on our results, using a simple theoretical model, this section discusses whether flow-driven turbulence can explain the observed line mass independent width~\citep{Arzoumanian2011A&A...529L...6A, Arzoumanian2019A&A...621A..42A}.}
Characteristics of an isothermal gas filament are often discussed using a static solution {first described by} \citet{Stodolkiewicz1963}:
\begin{equation}
\rho=\frac{\rho_{\rm c}}{\{1+(r/\sqrt{8}r_0)^2\}^2},
\label{eqs: density radial profile for isothermal cylinder}
\end{equation}
where $\rho_{\rm c}$ is the central density and $r_0\equiv c_{\rm s}/\sqrt{4\pi G\rho_{\rm c}}$, $\sqrt{8} r_0$ is a characteristic radius of the filament, where the density is one-quarter of the central density.
The line mass obtained from this solution gives the thermal critical line mass,
\begin{equation}
M_{\rm {line,cr,th}} \equiv \int_0^{\infty} 2 \pi \rho r d r=\frac{2 c_s^2}{G}.
\end{equation}
If the filament is confined by the external pressure $p_{\rm ext}$, the solution is truncated at radius $r_{\rm cut}$ where the gas pressure balances the external pressure; that is, $p_{\rm ext}=\rho(r_{\rm cut})\,c_{\rm s}^2$.
Since there is no equilibrium solution for supercritical filaments, such a solution has a subcritical line mass.
\citet{FischeraMartin2012A&A...542A..77F} derived an analytic expression of the column density structure of the truncated Stod\'olkiewicz solution.
The dashed curve of Figure \ref{FilW} plots the FWHM of the truncated Stod\'olkiewicz filament for $p_{\rm ext}/k_{\rm B}=2\times 10^4$ K cm$^{-3}$ as a function of line mass.
In this case, the central density uniquely determines the line mass.
Note that the conventional isothermal filament cannot maintain its width when the filament line mass is on the order or larger than the critical line mass.

The present simulations suggest that a fraction $f$ of the inflow velocity $v_{\rm flow}$ ($v_{\rm x}$ in our simulations) is converted into turbulent velocity dispersion $\Delta v^2=f^2\,v^2_{\rm flow}$.
To derive the FWHM width of the turbulent filaments, based on the results of our simulations we set $f=0.5$ in our model and replace the isothermal sound speed $c_{\rm s}$ and external pressure $p_{\rm ext}$ by the velocity dispersion $\Delta v$ and accretion flow ram pressure $\rho_{\rm ext}\,v_{\rm flow}^2$, respectively.
{We assume that a filament is in dynamical (quasi-)equilibrium at a given line mass, requiring that the transformation of inflow energy into the turbulence occurs in quasi-steady process.
This assumption is roughly justified, as the turbulent eddy turnover time ($\sim w_{\mathrm{FWHM}} / \Delta v = 0.1 \, \mathrm{pc} / 0.6 \, \mathrm{km \, s^{-1}} = 0.17 \, \mathrm{Myr}$) is shorter than the inflow timescale ($\sim 70 \, M_{\odot} \, \mathrm{pc^{-1}} / 2 \rho_{\rm sheet} H v_{x0} = 1.3 \, \mathrm{Myr}$).
In addition, observations suggest that filaments are in virial balance, where effective pressure forces (thermal, turbulent, and possibly magnetic fields) balance gravity~\citep{Arzoumanian2013A&A...553A.119A, Arzoumanian2021A&A...647A..78A, Shimajiri2023A&A...672A.133S}.
Interestingly, a similar conversion factor is reported in \citet{Clarke2017MNRAS.468.2489C, Clarke2018MNRAS.479.1722C, Clarke2020MNRAS.497.4390C}, who studied inflow onto a filament without a magnetic field.
One difference is that \citet{Clarke2017MNRAS.468.2489C, Clarke2018MNRAS.479.1722C, Clarke2020MNRAS.497.4390C} show this high conversion factor is possible only with inhomogeneous {and continuous} turbulent inflow (i.e., the turbulence in the inflow was {initially} added by hand), while our simulation shows the STORM mechanism can {naturally generate turbulence in the filament from homogeneous and non-turbulent inflows}.
{Although \citet{Clarke2020MNRAS.497.4390C} shows that a thermally supercritical filament can maintain its width due to inflow with supersonic turbulence in purely hydrodynamic simulations, some studies claim that accretion, in pure hydrodynamic and ideal MHD conditions, cannot result in filament stabilization~\citep{Heigl2020MNRAS.495..758H, Hacar2023ASPC..534..153H}.
In the simulations by \citet{Heigl2020MNRAS.495..758H}, the inflow is homogeneous, and thus, there is little asymmetry to drive the turbulence.
Consequently, the induced turbulence lacks a radial gradient.
Furthermore, our ideal MHD simulations without ambipolar diffusion show that the turbulent pressure is insufficient to support massive filaments against radial gravitational collapse~\citep[see also][]{Seifried2015MNRAS.452.2410S}.}
The critical finding in this article is that ambipolar diffusion can significantly sustain the filament width.
This is apparently puzzling and somewhat counter-intuitive because we tend to think that the diffusion erases the sharp structures and hence reduces the activity of turbulent motions.
The resolution of the puzzle is in the identification of the STORM mechanism explained in Section \ref{sec:Results}.
Here, we have newly identified a concrete mechanism to naturally drive non-decaying turbulence in a non-ideal MHD framework that is in contrast to the previous analyses in purely hydrodynamical cases.}
The characteristic radius $r_0$ and the effective critical line mass $M_{\rm {line,cr,eff}}$ are rewritten as
\begin{equation}
    r_0 = \frac{\Delta v_{\rm eff}}{\sqrt{4\pi G \rho_{\rm{c}}}},
\end{equation}
and
\begin{equation}
    M_{\rm {line,cr,eff}} = \frac{2 \Delta v_{\rm eff}^2}{G},
\end{equation}
respectively.
{{Where $\Delta v_{\rm eff} \equiv \sqrt{c_s^2 + \Delta v^2}$ is the effective velocity dispersion including thermal and turbulent velocity dispersion.}}
The truncation radius is determined so that the accretion ram pressure balances the local thermal pressure, i.e., $\rho(r_{\rm cut})\,c_{\rm s}^2=\rho_{\rm ext}\,v_{\rm flow}^2$.
Using Eq. (\ref{eqs: density radial profile for isothermal cylinder}), the truncation radius $r_{\rm cut}$ is given by
\begin{equation}
    r_{\rm cut} = \frac{\sqrt{8} \Delta v_{\rm eff}}{\sqrt{4\pi G \rho_{\rm{c}}}} \sqrt{\left( \frac{\rho_{\rm{c}}}{\rho_{\rm ext}} \right)^{1/2} \frac{c_s}{v_{\rm flow}} -1},
\end{equation}
The column density profile of a filament is given by \citet{FischeraMartin2012A&A...542A..77F}:
\begin{equation}
    N(x)=n_{\mathrm{c}} \sqrt{8} \frac{r_0}{c}\left\{\frac{u}{u^2+c}+\frac{1}{\sqrt{c}} \arctan \frac{u}{\sqrt{c}}\right\},
\end{equation}
{where $n_{\rm c}$ denotes the central number density, $x$ is an impact parameter in units of $r_{\rm cut}$, $c=1+x^2 r_{\mathrm{cut}}^2 / 8 r_0^2$, and $u=\sqrt{1-x^2} r_{\mathrm{cut}} /\left(\sqrt{8} r_0\right)$, respectively.
The FWHM is given by numerically solving the equation $N(x)=N(0)/2$ for $x$.}
The FWHMs for $v_{\rm flow}=1$ and $2$ km s$^{-1}$ and $n_{\rm ext}=10^3$ cm$^{-3}$ are plotted as functions of the line mass in Figure \ref{FilW}.
The plots show that, in contrast to the conventional case of an isolated filament~\citep[e.g.,][]{FischeraMartin2012A&A...542A..77F}, the filament formed by a converging flow shows {an FWHM of $\sim$ 0.1 pc even for largely supercritical line mass values}.
In addition, the {maximum width appears relatively insensitive to the inflow velocity} because increasing the inflow velocity reduces the $r_{\rm cut}$ but enlarges the $r_0$ by enhancing the $\Delta v$.
The balance of these two effects renders the filament width insensitive to the converging velocity value.
Note that the filaments reach larger line mass values for larger converging velocities.
{Our model yields an FWHM that is smaller by a factor of 2 compared to that obtained from the simulation (panel (c) in Figure \ref{fig:radialprofiles}), but it can still reproduce the overall trend.
{One possible explanation for this discrepancy is that our model assumes that the turbulence acts as an isotropic pressure, whereas in our simulations, turbulence is anisotropic and strong in the x-direction.
Since we measure the width along the x-direction, where the turbulence is the most developed, the FWHM predicted by our model is smaller than that measured in the simulation.}}
In conclusion, if the line mass of the filament is less than the effective critical line mass, $2\Delta v^2 / G$, the filament width will remain in the order of $\sim$ 0.1 pc (Figure \ref{FilW}).


The above model assumes isotropic accretion of gas onto the filament.
However, in reality, the accretion (and resulting turbulence) occurs one-dimensionally along the magnetic field lines.
In the direction perpendicular to the magnetic field, the magnetic pressure is responsible for the finite width.
In the type-O filament formation scenario, the magnetic pressure in the shocked sheet, where the filament forms, is as large as the accretion ram pressure, suggesting that we can expect a width similar to the present model in the direction across the magnetic field.
Using the parameter given by the simulation result, the width perpendicular to the magnetic field can be simply estimated as the effective Jeans length,
\begin{equation}
\begin{aligned}
    w_{\rm \perp} 
    & \sim 
    \left(v_A^2 + c_s^2 + \Delta v_z^2 \right)^{1/2} t_{\rm ff} \\ 
    & = \left(\frac{\bar{B}_{\rm fil}^2/4\pi \bar{\rho}_{\rm fil} + c_s^2 + \Delta v_z^2} {2\pi G \bar{\rho}_{\rm fil}} \right)^{1/2}
    = 0.036\ \mathrm{pc},
    \label{eqs: w_perp, effective jeans length}
\end{aligned}
\end{equation}
where, $\bar{\rho}_{\rm fil} = m_{\rm mol} \bar{n}_{\rm fil}$ ($\bar{n}_{\rm fil} = 1.1\times 10^5\,\rm{cm^{-3}}$), $\bar{B}_{\rm fil} = 87\ \mathrm{\mu G}$ and $\Delta v_z = 0.19\,\mathrm{km\,s^{-1}}$ denote the mean mass density, the mean magnetic field strength, and the velocity dispersion in z-direction in the filament given by our simulation {at $t$ = 1.15 Myr}.
{The width perpendicular to the magnetic field, $w_{\rm \perp}$, is close to the result of our simulation, $w_{\mathrm{FWHM,Pl,\perp}} = 0.033\,\mathrm{pc}$.}

\section{Summary} \label{sec: summary}
We investigated filament evolution under the influence of {the} slow-shock instability.
The major findings of this study are summarized below.
\begin{itemize}
  \item Ambipolar diffusion allows the post-shock gas to flow across the magnetic fields in the shock vicinity, forming dense blobs behind the concave points of the shock. The blobs transfer momentum that drives internal turbulence {(We name this mechanism \textit{``STORM"} and the blob ``hail" because the blobs resemble hails that are ejected like a hailstorm into the filament)}. This mechanism transforms the kinetic energy of the {converging} flows into anisotropic turbulence in the filament, providing an effective pressure that increases the filament width.
  \item The ideal MHD and non-magnetized models cannot develop sufficient turbulence, indicating that both magnetic field and ambipolar diffusion are necessary for maintaining a constant width of {thermally supercritical} filaments.
  \item {The size of each {hail}} (or fineness of structure in filaments) depends on the ambipolar diffusion coefficient. The strength of the turbulence driven by the {STORM} depends only on the {inflow} velocity and is insensitive to density, magnetic field, and ionization degree.
  \item {The STORM generates a turbulent pressure gradient inside the filament that prevents its radial collapse.}
  \item A filament of $\sim$ 100 $M_{\odot}$ pc$^{-1}$ formed under the influence of the {STORM} shows a width {compatible with the observations}.
  \item {We propose an analytical model showing} that the width of the turbulent filament confined by inflowing gas does not substantially depend on {inflow} velocity, however, the maximum line mass value depends on the inflow velocity.
\end{itemize}
{
{Although the STORM prevents its radial collapse of a filament of $\sim$ 100 $M_{\odot}$ pc$^{-1}$, the results of our simulation show that after $\sim$ 0.75 Myr, the accretion due to the gravity of the filament is not increasing sufficiently, leading to its collapse.
In contrast, some observations seem to show that higher line mass filaments have higher accretion rates indicating that filaments, regardless of their line mass remain in virial balance [roughly doubling their line mass in $\sim$ 1 Myr, \citet{Palmeirim2013}, \citet[][, PPVII]{Pineda2023ASPC..534..233P}] resulting in filaments with $>$ 100--500 $M_{\odot}$ pc$^{-1}$ and widths of 0.1 pc~\citep{andre2025arXiv250324316A}.
Some filaments might be more than 1 Myr old.
This may indicate that certain aspects of the evolution of filaments are not yet fully explored in the present analyses.}
Furthermore, it is possible that we overestimate or underestimate the STORM effect in this paper because we give steady inflows.
Actual filament formation and evolution are induced by flows driven in a shock-compressed layer~(see the left panel in Figure \ref{fig:inicon}).
{Moreover, it is realistic that such flows include density and velocity inhomogeneities.}
In a future paper, we will investigate the effect of STORM in the simulations of realistic filament formation induced by the primary shock that corresponds to the fast MHD shock, i.e., in the simulations including both fast and slow MHD shocks.}

\begin{acknowledgments}
We thank K. Tomida, who provided a new multi-grid module for implementing self-gravity in Athena++ code. We also thank P. Andr\'{e}, S. Takasao, K. Iwasaki, M.I.N. Kobayashi, and R. Kashiwagi for fruitful discussions. The numerical computations were carried out on the XC50 system at the Center for Computational Astrophysics (CfCA) of the National Astronomical Observatory of Japan. This work is supported by Grants-in-aid from the Ministry of Education, Culture, Sports, Science, and Technology (MEXT) of Japan (JP22J15861, 18H05436, 18H05437).
\end{acknowledgments}

\vspace{5mm}

\appendix

\section{Difference by Solvers} \label{sec: Difference by Solvers}
\begin{figure*}[ht!]
\epsscale{1.15}
\plotone{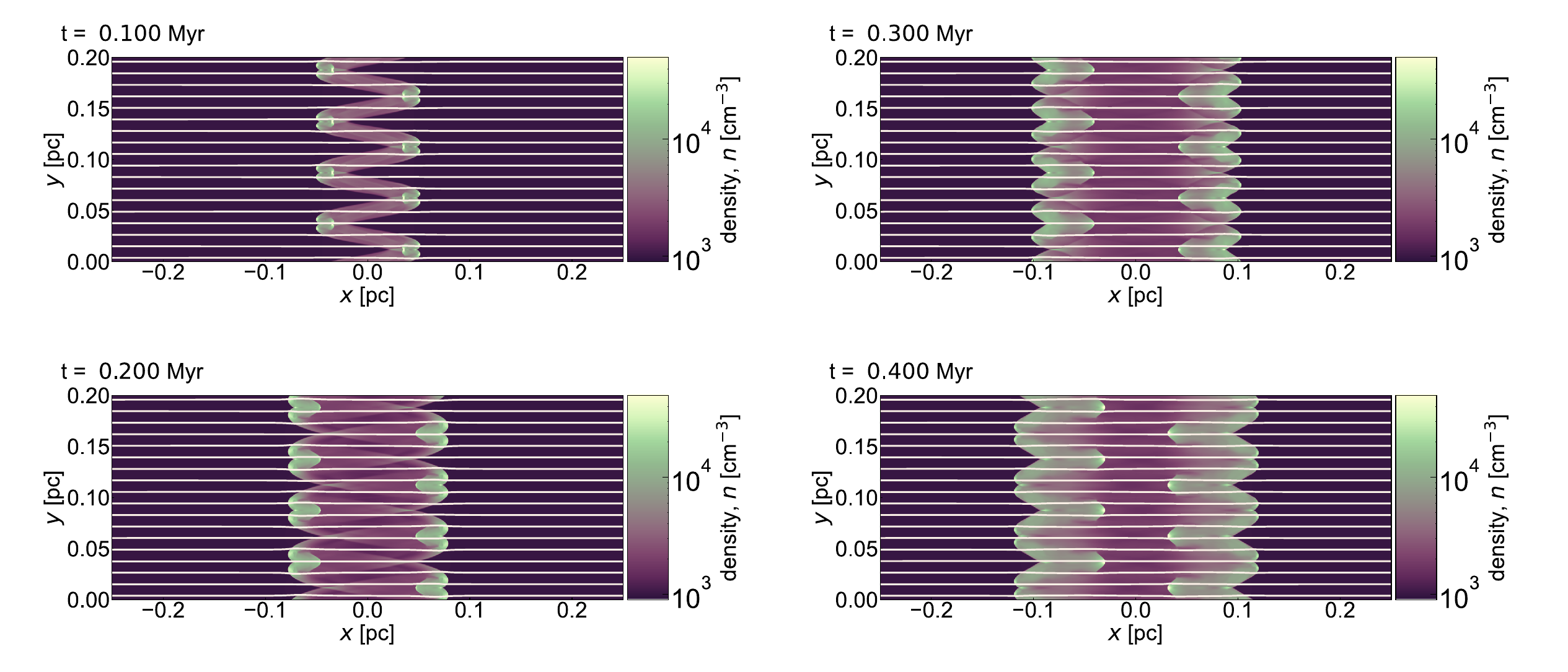}
\caption{\small{Similar to model SingleModeAD in Figure \ref{fig:2ddensity005} but usuig low-dissipation Harten-Lax-van Leer discontinuities (LHLLD) model.
{As the LHLLD and Roe solver obtain very similar results in the simulations with ambipolar diffusion.}
\label{fig:2ddensity005lhlld}
}}
\end{figure*}
In the ideal MHD case, the simulation of double-slow mode shocks fails when using the Roe solver.
Therefore, to simulate the nonlinear evolution of SSI in the ideal MHD case, we employ the low-dissipation Harten-Lax-van Leer discontinuities (LHLLD) model.
As the LHLLD and Roe solver obtain very similar results in the simulations with ambipolar diffusion, the comparison in Figure \ref{fig:2ddensity005} in \S \ref{subsec: Two-dimensional simulations w/o self-gravity} is justified, although we use different MHD solvers.
{Furthermore, this consistent result indicates that the carbuncle phenomenon does not affect the results in Figure \ref{fig:2ddensity005} since LHLLD solver is designed to prevent the carbuncle phenomenon.}

\bibliography{ms}{}
\bibliographystyle{aasjournal}



\end{document}